\documentclass[numberedappendix]{emulateapj}

\slugcomment{ApJ, to be submitted; \today}

\newcommand{\sT}{\sigma_{\rm T}}

\newcommand{\e}{\epsilon}
\newcommand{\g}{\gamma}

\newcommand{\psim}{\lower.5ex\hbox{$\; \buildrel \propto \over\sim \;$}}
\newcommand{\lbar}{\lower.0ex\hbox{$\; \buildrel
{\lower0.0ex \hbox{-}} \over\lambda  \;$}}

\newcommand{\cm}{\mathrm{cm}}

\newcommand{\erg}{\mathrm{erg}}
\newcommand{\eV}{\mathrm{eV}}

\newcommand{\TeV}{\mathrm{TeV}}
\newcommand{\s}{\mathrm{s}}

\newcommand{\Hz}{\mathrm{Hz}}

\newcommand{\kpc}{\mathrm{kpc}}

\newcommand{\yr}{\mathrm{yr}}

\newcommand{\Gauss}{\mathrm{G}}
\newcommand{\Kelvin}{\mathrm{K}}

\newcommand{\rxj}{RX~J1713.7-3946}
\newcommand{\fermi}{{\em Fermi}}

\shorttitle{Electron Evolution in Supernova Remnants}
\shortauthors{Finke \& Dermer}

\begin{document}
\title {Cosmic-ray Electron Evolution in the Supernova Remnant \rxj~}

\author{Justin D. Finke and Charles D. Dermer}

\affil{U.S.\ Naval Research Laboratory, Code 7653, 4555 Overlook Ave. SW,
        Washington, DC,
        20375-5352\\
}

\email{justin.finke@nrl.navy.mil}

\begin{abstract}

A simple formalism to describe nonthermal electron acceleration,
evolution, and radiation in supernova remnants (SNRs) is presented.
The electron continuity equation is analytically solved assuming that
the nonthermal electron injection power is proportional to the rate at
which the kinetic energy of matter swept up in an adiabatically
expanding SNR shell.  We apply this model to \fermi\ and HESS data
from the SNR \rxj, and find that a one-zone leptonic model with
Compton-scattered cosmic microwave background (CMB) and interstellar
infrared photons has difficulty providing a good fit to its spectral
energy distribution, provided the source is at a distance $\sim 1\
\kpc$ from the Earth.  However, the inclusion of multiple zones, as
hinted at by recent {\em Chandra} observations, does provide a good
fit, but requires a second zone of compact knots with magnetic fields
$B\sim 16\ \mu$G, comparable to shock-compressed fields found in the
bulk of the remnant.

\end{abstract}

\keywords{supernova remnants --- acceleration of particles --- 
radiation mechanisms:  nonthermal --- shockwaves --- gamma rays:  theory}
	  
% changes from v09:
%
% removed ``easily''
% added references (Lagage, Berezinskii, etc. in intro)
% minor rewording.	  
% end of page 1/beginning of page 2 (emulateapj format), added
% bit about synch/Compton-scattered emission from dense regions.
% page 2, first column:  we do not neglect the bremsstrahlung process.
% same place:  particle acceleration rate doesn't decline for $t<t_s$.
% intro to section 2 (before 2.1) unnecessary?  Seems it is repeated 
% in section 2.1.  
% before equation (30):  I do not take k_ad = 2/5.  There is now no 
% 2/5 in equations (5), (6), or (8).  
% Using 2/5 means the solution will not be analytic.  

% changes from v10 (28 Sept. 2011 version) by JDF:
% changes \S to section.  ApJ no longer uses \S symbol.
% changed f(t) and solutions based on it.
% Figures updated with new calculation, table 1 updated with new fit.
% removed ``NASA Guest Investigator Grants'' from the acknowledgements.
% Our Fermi GI grant on SNR wasn't accepted.
% Some edits to the text around equation 10.

\section{Introduction}
\label{intro}

Acceleration of particles at SNR shocks is considered the leading
mechanism for the production of cosmic-ray protons and ions from
energies of $\approx 1$ GeV/nucleon up to the knee of the cosmic-ray
spectrum \citep[e.g.,][]{lagage83,blandford87,berez90}.  As the
expanding SNR shell sweeps up matter from the surrounding
circumstellar medium (CSM), a pair of shocks is formed, with the
leading forward shock sweeping up and accelerating CSM material, and a
reverse shock braking the metal-rich SNR ejecta.  A few select
particles gain energy as they randomly diffuse back and forth across
each of the two shock fronts, while convecting downstream into the
shocked fluid.  In the test particle theory of first-order Fermi
acceleration, this leads to a nonthermal power-law particle
distribution in momentum with number spectral near $-2$ for a
compression ratio near 4.  The spectrum that results after folding in
the effects of diffusive escape from the SNR into intergalactic space,
and from the disk of the Galaxy into the halo, is in reasonable
agreement with the measured cosmic-ray spectrum
\citep[e.g.,][]{jones91,kirk94,hillas05}.

Charged cosmic rays are deflected by Galactic magnetic fields during
transport, so their direction does not point back to the original
production site.  This is the principal reason that the sources of
Galactic cosmic rays remain elusive a century after their
discovery. Given the extreme difficulty in detecting neutrinos
\citep{yuan11_neutrino}, electromagnetic signatures of cosmic rays
offer at present the best opportunity to identify the sources of
cosmic rays and settle the question of their origin 
\citep[e.g.,][]{ginzburg64,gaisser90,drury94,reynolds08}.

With speeds reaching $10^4$ km s$^{-1}$ or more, SNR shocks are likely
candidates for accelerating electrons and nucleons to high energies.
The polarization and spectral properties of the smooth broadband
nonthermal radio through X-ray emission is almost certainly electron
synchrotron radiation, though the optical, UV and X-ray spectra can
additionally reveal strong line signatures from shock-heated shell
material \citep{slane02,badenes06}.  Several radiative mechanisms can
be responsible for $\gamma$-ray emission. Energetic leptons emit
$\gamma$ rays through Compton scattering of the ambient radiation
fields, principally the cosmic microwave background (CMB) and ambient
stellar and IR fields, but also by scattering photons of the the
synchrotron field.  Electrons make nonthermal bremsstrahlung $\gamma$
rays when colliding with target gas and dust particles in the
circumstellar medium (CSM) or in the shocked shell material.  Nuclear
collisions of hadrons make pion-decay $\gamma$ rays when cosmic-ray
protons and ion interact with that same matter.  Spectral and
morphological differences are expected between a leptonic and hadronic
origin. Most well known is the prediction of the $\pi^0$ decay feature
peaking at 70 MeV in a photon spectrum \citep{ginzburg64,hayakawa69}
that results from hadronic processes. For an electron injection
spectrum softer than number index $q=2$ (where the injected electrons
are injected with spectrum $Q(\g)\propto \g^{-q}$), as expected in the
test-particle limit for strong nonrelativistic shocks in a hydrogen
medium, the Compton-scattered radiation spectrum is much harder than
the electron bremsstrahlung $\gamma$-ray spectrum.  For a consistent
explanation, the synchrotron spectrum must also be compatible with the
same electron distribution that makes the $\gamma$ rays.

Differences between morphological features can help discriminate
between leptonic and hadronic origins of $\g$ rays in SNRs.  Emission
of $\g$ rays from bremsstrahlung and proton interactions is expected
to be enhanced in the vicinity of dense molecular clouds, not only due
to the denser target material, but also due to the increased amount of
energy dissipated at the shock front
\citep[e.g.,][]{bykov00,aharonian96,gabici09}. Almost all SNRs
associated with {\em Fermi} sources exhibit OH maser emission from
SNR/molecular cloud interactions \citep{hewitt09}.  By
contrast, synchrotron X-rays and TeV $\gamma$-rays from
Compton-scattered CMB photons require only high-energy electrons and a
magnetic field, and in principle could be found in regions of small
gas density.

Recent \fermi\ Large Area Telescope (LAT) observations have provided a
wealth of data on SNRs. The First \fermi\ Catalog of Gamma Ray Sources
\citep[1FGL;][]{abdo10_1fgl} lists 41 associations with sources in
Green's SNR catalog \citep{green09}. Morphological similarities allow
definite identifications to be made in 3 cases: 1FGL J1856.1+0122 with
W44 \citep{abdo10_w44}; 1FGL J1922.9+1411 with W51C
\citep{abdo09_w51c}; and 1FGL J0617.2+2233 with IC 443
\citep{abdo10_ic443}. In the second {\em Fermi}-LAT catalog
\citep[2FGL;][]{abdo11_2fgl}, 3 additional SNR identifications are
reported besides a total of 62 associations with SNRs and pulsar wind
nebulae.  More recently, \rxj\ has been added to this list
\citep{abdo11_rxj}.  The \fermi\ data provide strong evidence that
$\g$-ray emission is made by accelerated particles in the vicinity of
these objects. Yet it is not conclusive that the SNRs are accelerating
the particles, as pre-existing cosmic rays compressed by the
outflowing remnant could make the emission \citep{uchiyama10}.

There have been numerous hadronic and leptonic models produced to
explain the particle acceleration and $\g$-ray emission from SNRs,
with varying degrees of complexity
\citep[e.g.,][]{sturner97,baring99,zirak07,lee08}.  Here we focus on a
simple model for electron acceleration at the forward SNR shock and
study the evolving distribution with the addition of radiative and
adiabatic losses \citep[however, note that acceleration in the
reverse shock could be substantial;][]{telez12}.  
The goal is to explain the \rxj\ spectrum with a
purely leptonic model involving Compton scattering of diffuse target
photons and a possible small contribution from electron
bremsstrahlung.  As the SNR expands into the CSM and decelerates due
to the addition of the swept-up matter, the particle injection power
initially increases in the free-expansion phase, and subsequently
declines in the Sedov phase.  The time-dependent injection efficiency
is very model dependent, and ultimately rests on microphysical plasma
processes.  Here we normalize the power of injected nonthermal
electrons to the swept-up power; other normalizations could employ an
injection efficiency proportional to the rate at which particle mass
is swept up, or an efficiency dependent on shock speed and compression
ratio.
%We consider modifications of the energy injection rate.

%This
%variability in injection can lead to signatures in the nonthermal
%electron distribution, which in turn are observable in the SED, and
%this is the focus of this paper.

In Section \ref{formalism} we describe our model. In Section
\ref{rxj_section}, we use it to fit the multiwavelength SED of the SNR
\rxj\ (G~347.3$-$0.5), and show that a single-zone model is incapable
of fitting the spectrum. A model with a second zone of emission
regions consisting of compact knots is shown to give a good fit to the
SED.  We conclude with a summary and discussion in Section
\ref{conclusion}.

\section{Formalism}
\label{formalism}

We make a number of common simplifying assumptions.  A
spherically-symmetric supernova explosion finds itself in a
homogeneous surrounding CSM with constant number density
$n_{CSM}$. The explosion is approximated by an expanding shell of
matter that sweeps up CSM material. The inclusion of the swept-up mass
controls the dynamics of the shell, and the system proceeds to channel
directed kinetic energy into internal kinetic energy of the shocked
matter.  The injection rate changes abruptly at the Sedov age.  We
purposely keep the transition between the pre-Sedov (i.e., free
expansion) and Sedov phases discrete to highlight interesting
injection effects, keeping in mind that a more detailed treatment
would have a smooth pre-Sedov and post-Sedov transition.

\subsection{SNR Dynamics}
\label{dynamics}

Here we describe a simple formalism for the dynamics of SNRs.  A
detailed hydrodynamic description is given by \citet{truelove99}.  The
kinetic energy, $E$, of the remnant and swept up matter is conserved
in an adiabatic blast wave, so that
\begin{eqnarray}
\label{energy}
E = \frac{1}{2} M_0v_0^2 = 
\frac{1}{2}\left[ M_0 + \frac{4\pi}{3}m_p n_{CSM} r(t)^3 \right]v^2(t) \ ,
\end{eqnarray}
where $M_0$ and $v_0$ are the initial remnant's mass and speed,
respectively, $m_p$ is the proton mass, and $r(t)$ and $v(t) = dr/dt$
are respectively the radius and speed of the SNR as a function of time
$t$.  For simplicity, the CSM is assumed to be composed entirely of
hydrogen.  The Sedov radius, $r_s$, is defined as the radius where the
mass of the swept-up CSM matter, $4\pi m_p n_{CSM} r(t)^3/3$, is equal
to the mass of the initial explosion, $M_0$, i.e.,
\begin{eqnarray}
\label{rsedov}
r_s \equiv \left[ \frac{3M_0}{4\pi m_p n_{CSM}}\right]^{1/3}\ .
\end{eqnarray}
Using Equation (\ref{rsedov}), Equation (\ref{energy}) can be rewritten 
\begin{eqnarray}
\label{energy2}
v_0^2 = 
\left[ 1 + \left(\frac{r}{r_s}\right)^3 \right]\left(\frac{dr}{dt}\right)^2\ ,
\end{eqnarray}
or
\begin{eqnarray}
\label{exactDE}
v(t) = \frac{dr}{dt} = \frac{v_0}{\left[1 + (r/r_s)^3\right]^{1/2}}\ .
\end{eqnarray}
This can be solved in the limit $r \ll r_s$ , giving
\begin{equation}
\label{rt1}
r(t) = v_0 t\ ,
\end{equation}
which is known as the free expansion phase of the remnant. 
In the Sedov phase, $r \gg r_s$, giving the well-known behavior 
\begin{equation}
\label{rt2}
r(t) = v_0 t_s \left( \frac{5t}{2t_s}\right)^{2/5}\ .
\end{equation}
Here the Sedov time is defined as
\begin{eqnarray}
\label{tsedov}
t_s \equiv \frac{r_s}{v_0}\ . 
\end{eqnarray}
The solutions (\ref{rt1}) and (\ref{rt2}) intersect when $t=1.84t_s$.
The speed of the remnant in these limits is 
\begin{equation}
\label{speed_sedov}
v(t) = \left\{ \begin{array}{ll}
    v_0 & r \ll r_s \\
    v_0\left(5t/2t_s\right)^{-3/5} & r \gg r_s 
  \end{array}
  \right. \ .
\end{equation}
The solutions for $r\ll r_s$ and $r\gg r_s$ intersect when
$t=0.40t_s$.  The rate at which kinetic energy is swept up from the
surrounding CSM is
\begin{eqnarray}
\label{SNRpower}
%\frac{dE}{dt} \approx \frac{1}{2}\frac{dM}{dt}v^2 = 
%\frac{1}{2}(4\pi n_{CSM} m_p r^2 v)\ v^2 = 2\pi r^2 n_{CSM} m_p v^3\ .
\frac{dE}{dt} = 2\pi r^2 n_{CSM} m_p v^3\ .
\end{eqnarray}

\begin{figure*}
\vspace{2.0mm}
\epsscale{0.8}
\plottwo{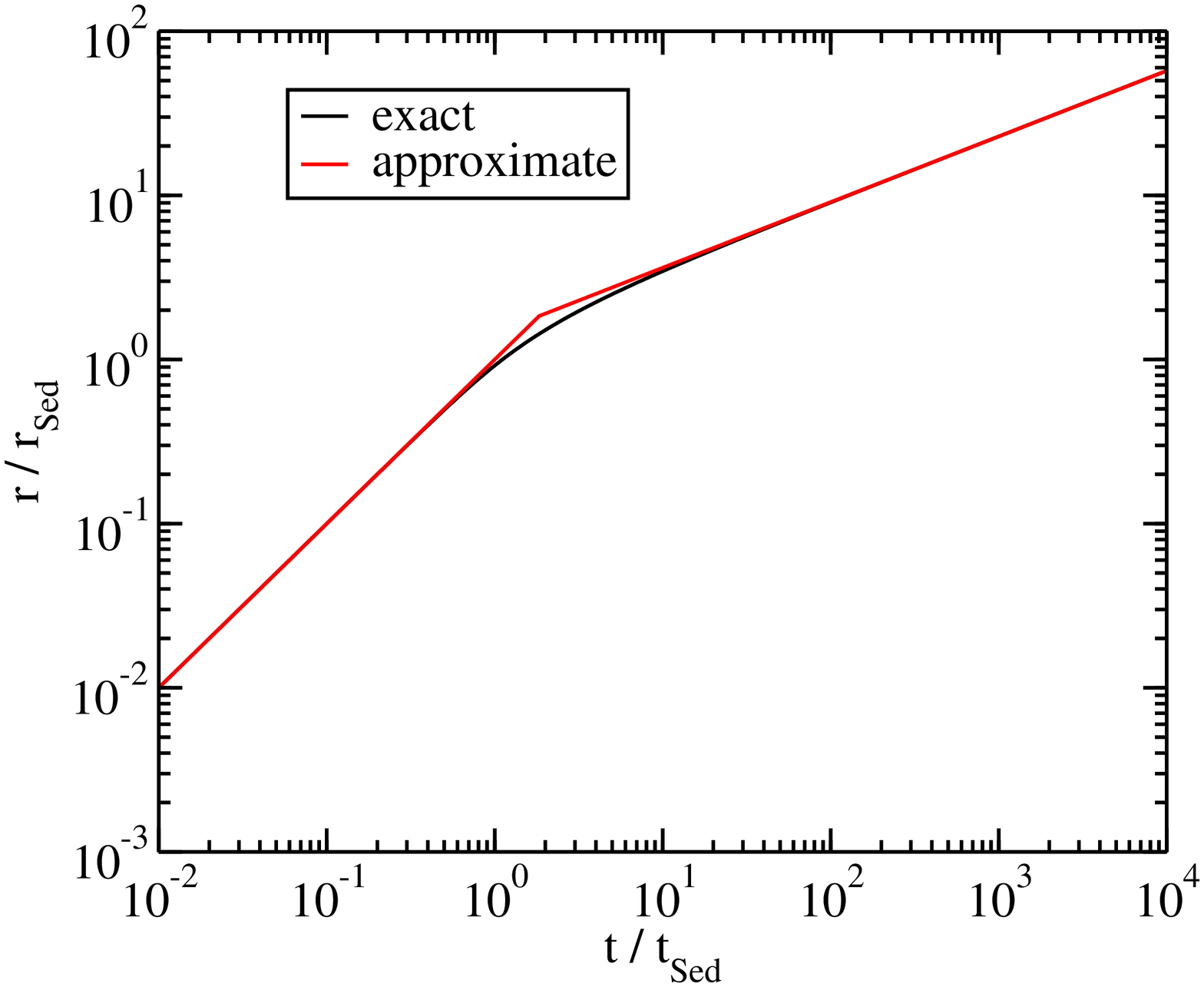}{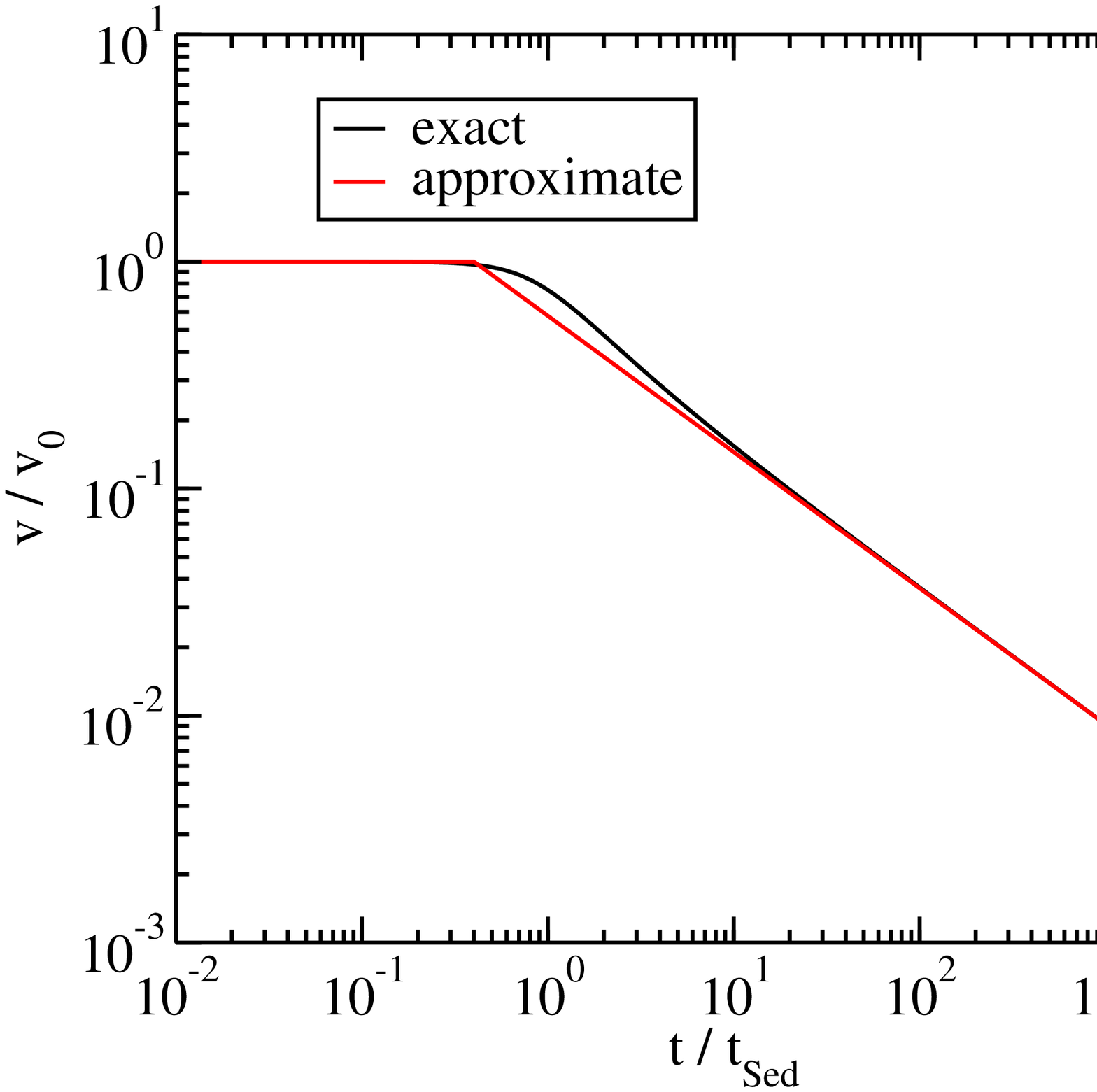}
\caption{Evolution of SNR radius and speed using the exact and approximate 
expressions.
}
\label{param_evo}
\vspace{3mm}
\end{figure*}

In Figure \ref{param_evo} we plot the radius and speed of the remnant 
for the exact expression, Equation (\ref{exactDE}), and the 
approximate expressions, Equations (\ref{rt1}), (\ref{rt2}), and 
(\ref{speed_sedov}).  As can be seen, the approximate form reproduces 
the exact behavior quite well.

%\clearpage

\subsection{Particle Acceleration}
\label{part_accel}
As the SNR shock expands into the CSM, particles will be accelerated
at the forward shock, to which we restrict our treatment.  
We assume that the injected kinetic
energy of the nonthermal particle distribution is some fraction $\eta$
of the swept-up kinetic energy. This energy is swept into the shocked
fluid and allows us to normalize the the nonthermal injection function, 
$Q(\g,t)$  by the relation
\begin{eqnarray}
\label{part_power}
m c^2 \int^{\g_2}_{\g_1} d\g\ \g\ Q(\g, t) = \eta 2\pi r^2 n_{CSM} m_p v^3\ ,
\end{eqnarray}
where $\g$ is the particle's Lorentz factor and $m$ is the particle's
mass.  In order not to sweep in more energy than was originally
available, $\eta$ is restricted to be $\ll 1$, and the treatment is
restricted to an adiabatic blast wave.

\begin{figure}
\vspace{2.2mm}
\epsscale{0.7}
\plotone{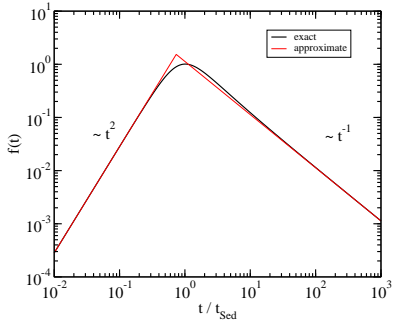}
\caption{The exact (Equation [\ref{f_exact}]) and approximate
(Equation [\ref{f_approx}]) expressions for $f(t)$.
}
\label{power_norm}
\vspace{2.0mm}
\end{figure}

Assuming that the injected accelerated particle distribution function
can be described by a power law, then
\begin{equation}
Q(\g, t) = Q_0(t)\ \g^{-q} H(\g;\g_1,\g_2)\ ,
\end{equation}
where $H(x;a,b) = 1$ if $a \leq x\leq b$, and $H(x;a,b) = 0$
otherwise.  Equation (\ref{part_power}) can be integrated to give
$$Q_0(t)= \frac{r^2 v^3\ 2\pi n_{CSM} m_p \ \eta}{m c^2}$$
\begin{eqnarray}
\label{Q0}
\times \left\{ \begin{array}{ll}
(q-2)( \g_1^{2-q} - \g_2^{2-q})^{-1} & {\rm for~} q \ne 2 \\ \\
\left[\ln(\g_2/\g_1)\right]^{-1} & {\rm for~} q = 2 
\end{array}
\right.\ .
\end{eqnarray}
Equation (\ref{exactDE}) can be inserted into Equation (\ref{Q0}) and
used to write an approximate expression for $Q(\g,t)$,
\begin{equation}
Q(\g,t) = Kf(t) \g^{-q} 
\end{equation}
where 
$$K \equiv \frac{v_0^5 t_s^2\ 2\pi n_{CSM} m_p\ \eta}{m c^2}$$
\begin{equation}
\label{Kconst}
 \times
\left\{ \begin{array}{ll}
(q-2)( \g_1^{2-q} - \g_2^{2-q})^{-1} & q \ne 2 \\
\left[\ln(\g_2/\g_1)\right]^{-1} & q = 2 
\end{array}
\right.\ 
\end{equation}
and
\begin{equation}
\label{f_exact}
f(t) = \frac{ (r/r_s)^2 } { \left[ 1 + (r/r_s)^3 \right]^{3/2} }\ .
\end{equation}
The approximation
\begin{equation}
\label{f_approx}
f(t) \approx \left\{ \begin{array}{ll}
  (t/t_s)^2 & t < Ct_{s} \\
  (5t/2t_s)^{-1} & t > Ct_{s}
\end{array}
\right.\ .
\end{equation}
is in accord with the asymptotes for $r(t)$ and $v(t)$ from Section
\ref{dynamics}.  The division of the two branches of the approximation
at $Ct_s$ where $C\approx 0.74$ was chosen to produce a continuous
function.  The exact expression for $f(t)$, Equation (\ref{f_exact})
is compared with the approximation from Equation (\ref{f_approx}) in
Figure \ref{power_norm}.  As can be seen, the approximation is quite
good, with small discrepancies around $t\approx t_s$.

Using the power in the swept-up CSM to normalize the rate of particle
acceleration, as described above, is also standard for GRBs
\citep[e.g.,][]{chiang99} but differs from what has been done in the
past for SNRs. Earlier normalizations related the number of
accelerated electrons to the number of electrons swept into the
expanding blast wave \citep[e.g.,][]{sturner97,reynolds98,baring99},
which leads to a considerably different time-dependence in the Sedov
phase for the accelerated electrons ($Q(\g,t)\propto t^{1/5}$ instead
of $Q(\g,t)\propto t^{-1}$)\footnote{Note the typographical errors in
\citet{reynolds98} and \citet{baring99} who have $Q(\g,t)\propto
t^{-1/5}$.}.

The maximum particle energy can be calculated by equating the
acceleration time with the radiative loss time or the age of the
remnant \citep{reynolds98}.  To keep the treatment analytic, here we
assume $\g_{max}$ is constant in time.

\subsection{Particle Evolution}
\label{particle_evo}

When $Q(\g,t)$ has been determined (Section \ref{part_accel}), the
evolution of the particle distribution, $N(\g;t)$ can be found by
solving the continuity equation,
\begin{eqnarray}
\label{cont_eqn}
\frac{\partial N}{\partial t} + 
\frac{\partial}{\partial\g}\left[ \dot\g\ N(\g;t)\right] + 
\frac{N(\g;t)}{t_{esc}(\g,t)} = Q(\g,t)\ ,
\end{eqnarray}
where $t_{esc}$ is the escape timescale and $\dot\g$ is the cooling
rate.  Analytic solutions to the particle continuity equation
(\ref{cont_eqn}) are discussed in \citet{kardashev62,blumen70}; and
\citet[][Appendix C]{dermer09_book}.

\subsubsection{Solution with Radiative Losses}
\label{radiativelosses}

For electrons, escape timescales can be long and hence will be
neglected.  The electron Lorentz factor $\bar \gamma$ above which
synchrotron losses dominate bremsstrahlung losses is $\bar\gamma
\approx 5\times10^4\ (n_{CSM}/{\rm cm}^{-3})(10\ \mu {\rm G}/B)^2$,
but the corresponding timescale for energy loss is $\approx 5\times
10^7$ yrs.  This is much longer than the age of the remnant even for
very dense target material, so bremsstrahlung losses can be safely
neglected.  Thus we assume electron energy losses are dominated by
radiative losses from synchrotron and Thomson scattering of CMB
photons, so that
\begin{eqnarray}
-\dot\g = \nu\ \g^2
\end{eqnarray}
where
\begin{eqnarray}
\label{nu_cool}
\nu = \frac{4c \sT [B^2/(8\pi) + u_{CMB}] }{3m_e c^2}\ 
\\ \nonumber
= 1.34\times 10^{-20}~{\rm s}^{-1} [1+(B/3.23\ {\mu}\Gauss)^2]  \; ,
\end{eqnarray}
$B$ is the magnetic field in the remnant, and $u_{CMB} =
4.13\times10^{-13}\ \erg\ \cm^{-3}$ is the energy density of the CMB
at the present epoch.  Klein-Nishina effects should be of little
importance to the evolution of the electron spectrum as long as the
synchrotron losses dominate over Compton losses, which will be the
case for $B\ga3\ \mu \Gauss$.  In this situation, the continuity
equation has the solution
\begin{eqnarray}
\label{integ_soln}
\g^2 N_e(\g;t) = K \int_{t_{min}}^t\ dt_i\ \g_i^{2-q}\ f(t_i)\ 
\end{eqnarray}
\citep[see Appendix \ref{appendix_radonly_soln};
also][]{kardashev62,blumen70,dermer98}.  Here
\begin{equation}
\g_i = \frac{1}{\g^{-1} - \nu(t - t_i)}\ ,
\end{equation}
and 
\begin{equation}
t_{min} = \max[0, t - \nu^{-1}(\g^{-1} - \g_2^{-1})]\ .
\end{equation}

It is instructive to look at the case $q=2$, where the integral in
Equation (\ref{integ_soln}) can be performed analytically.  For
$t<Ct_s$,
\begin{equation}
\g^2 N_e(\g;t) = \frac{K t_s}{3}
\left[ \left( \frac{t}{t_s}\right)^3 - \left( \frac{t_{min}}{t_s} \right)^3 \right]\ ,
\end{equation}
while for $t>Ct_s$ and $t_{min} < Ct_s$, 
\begin{equation}
\g^2 N_e(\g;t) = K t_s \left[ 
\frac{1}{3}\left(C^3 - \left(\frac{t_{min}}{t_s}\right)^3\right) + 
\frac{2}{5}\ln\left(\frac{t}{C t_s}\right) \right]\ .
\end{equation}
For $t>Ct_s$ and $t_{min}>Ct_s$, 
\begin{equation}
\g^2 N_e(\g;t) = \frac{2K t_s}{5} \ln\left(\frac{t}{t_{min} }\right) \ .
\end{equation}

Now we examine the asymptotes for this solution, starting with the
case where $t<Ct_s$.  In the limit $t \ll (\nu\g)^{-1}$ or $\g \ll
(\nu t)^{-1}$, $t_{min} \rightarrow 0$, and
\begin{eqnarray}
\label{gNg_cool1}
\g^2 N_e(\g;t) \approx K\ \frac{t^3}{3 t_s^2} \propto t^3\ \g^0\ .
\end{eqnarray}
For $t \gg (\nu\g)^{-1}$ or $\g \gg (\nu t)^{-1}$, 
\begin{equation}
\label{gNg_cool2}
\g^2 N_e(\g;t) \approx \frac{K}{3\nu\g}\left(\frac{t}{t_s}\right)^2\ 
\propto t^2 \g^{-1}.
\end{equation}
Thus, at low $\g$ ($\g \ll (\nu t)^{-1}$), the electron distribution
will have the power-law injection index (i.e., $N_e(\g;t)\propto
\g^{-q}$), and at high $\g$ ($\g \gg (\nu t)^{-1}$), the electron
distribution will be $N_e(\g;t) \propto \g^{-q+1}$, as seen in
Equations (\ref{gNg_cool1}) and (\ref{gNg_cool2}) and as expected for
a cooling distribution.  Since the electron distribution increases at
different rates in the different regimes, however, an inflection will
open up in the normalization of the distribution in these two regimes.

\begin{figure}
\vspace{2.0mm}
\epsscale{1.0}
\plotone{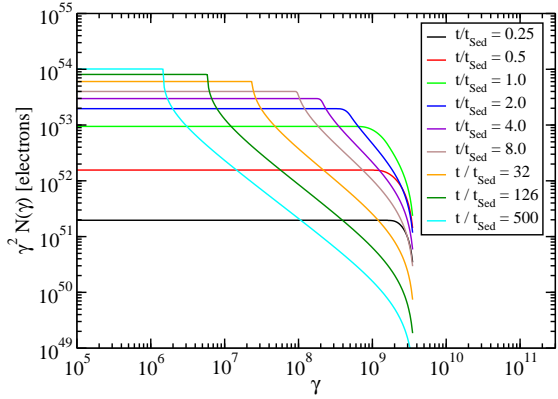}
\caption{Evolution of electron distribution with radiative losses only
for $q=2$.  Parameters are given in Table \ref{param_table1}.
}
\label{electron1}
\vspace{2.0mm}
\end{figure}

\begin{figure}
\vspace{2.0mm}
\epsscale{1.0}
\plotone{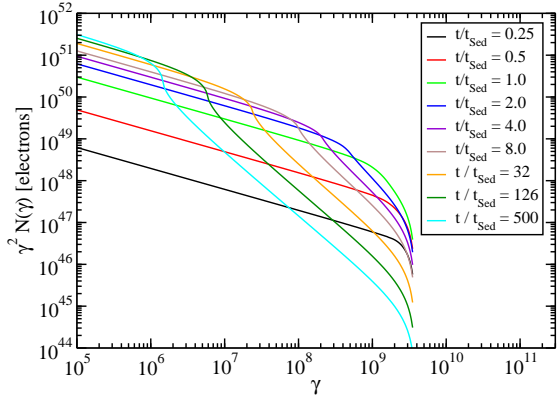}
\caption{Same as Figure \ref{electron1} for $q=2.5$.  Parameters are 
given in Table \ref{param_table1}.
}
\label{electron2}
\vspace{2.0mm}
\end{figure}

This effect becomes more pronounced for $t>Ct_s$.  In this case, if
$\g \ll (\nu t)^{-1}$,
\begin{eqnarray}
\label{g2N_1}
\g^2 N_e(\g;t) \approx K t_s \left[ \frac{1}{3} + 
\frac{2}{5}\ln\left(\frac{t}{Ct_s}\right) \right] \propto \g^0\ \ln(t/Ct_s)\ .
\end{eqnarray}
For $\g \gg (\nu t)^{-1}$ and $\g>\nu^{-1}(t-t_s)^{-1}$, 
\begin{eqnarray}
\label{g2N_2}
\g^2 N_e(\g;t) \approx \frac{K t_s}{\nu\g t} \propto t^{-1}\ \g^{-1}\ .
\end{eqnarray}
Thus, at low $\g$ ($\g \ll (\nu t)^{-1}$), the electron distribution
will again have the same power-law index with as the injection term,
just as with $t<Ct_s$, and the overall normalization will increase
with time, although more slowly than at $t<Ct_s$.  At large $\g$, the
electron distribution will be $N_e(\g;t)\propto \g^{-q+1}$, but the
normalization will be {\em decreasing} with time.  This will lead to
an increasingly large gap in normalization of the cooled ($\g < (\nu
t)^{-1}$) and uncooled ($\g > (\nu t)^{-1}$) electrons as time
increases.  This behavior can be seen in Fig.\
\ref{electron1}. Similar behavior can be found for $q \neq 2$, as
demonstrated in Fig.\ \ref{electron2}, where the integral in Equation
(\ref{integ_soln}) is performed numerically.  The parameters used 
in these calculations can be found in Table \ref{param_table1}. 

%\clearpage
%\begin{deluxetable*}{lc}
\begin{deluxetable}{lcccccc}
\tabletypesize{\scriptsize}
\tablecaption{
Test model parameters.
}
\tablewidth{0pt}
\tablehead{
\colhead{Parameter} &
\colhead{Symbol} &
\colhead{Model } &
}
\startdata
Blast Energy [erg]& $E$ & $1.0\times10^{51}$ \\
Initial Mass [$M_\odot$] & $M_0$  & 1.6\\
Initial Velocity [$\cm\ \s^{-1}$] & $v_0$  & $8.0\times10^8$ \\
ICM density [$\cm^{-3}$] & $n_{ICM}$  & 1.0 \\
Sedov time [yr] & $t_{s}$ & 303 \\
\hline
Magnetic field [$\mu$G] & $B$ & 10 \\
Cooling Constant [$\s^{-1}$] & $\nu$ & $1.4\times10^{-19}$ \\
\hline
Low energy electron cutoff & $\g_1$ & 10 \\ 
High energy electron cutoff & $\g_2$ & $3.1\times10^9$ \\
%Injection spectral index & $q$ & $2.1$ & $2.1$ & $2.1$ & $2.1$ & $2.1$ \\
Electron acceleration efficiency & $\eta_e$ & $10^{-4}$ \\
\enddata
\label{param_table1}
\vspace{2.0mm}
%\end{deluxetable*}
\end{deluxetable}
%\clearpage

%\clearpage

%\clearpage

\subsubsection{Solution with Adiabatic Losses}
\label{adiabaticlosses}

As the remnant expands, particles lose energy due to adiabatic losses,
since they are trapped in the expanding SNR.  The loss rate from this
process is
\begin{eqnarray}
-\dot\g = \frac{\g}{3}\frac{dV}{dt}\frac{1}{V} = \frac{k_{ad}\g}{t}
\end{eqnarray}
\citep[e.g.,][]{gould75} where the remnant's volume is $V\propto
r(t)^3$.  Details of the expansion control the adiabatic coefficient
$k_{ad}$, and in principle one can imagine a shell that contracts in
width while expanding outward to give no volume change, so
$k_{ad}\rightarrow 0$ for this peculiar system \citep[see, e.g.][for a
hydrodynamic description]{truelove99}.  In our approximations,
$k_{ad}=1$ for $r\ll r_s$, and $k_{ad}=2/5$ for $r\gg r_s$, given
$r(t)$ and $v(t)$ from Equations (\ref{rt1}), (\ref{rt2}), and
(\ref{speed_sedov}).  The results are only weakly dependent on
the value of $k_{ad}$, and so here we will assume the same $k_{ad}$ in
both regimes, eventually taking $k_{ad}=1$ for simplicity.  See
\citet{reynolds08} for a discussion on the dependence of the shock
radius with time for core collapse and type Ia progenitors, which has
implications for $k_{ad}$.

If adiabatic cooling dominates and radiative cooling is negligible,
then the solution to the continuity Equation gives
\begin{eqnarray}
N(\g;t) = Kt\int^t_{t_{min}} \frac{dt_i}{t_i}\ 
\left( \frac{\g t^{k_{ad}}}{t_i^{k_{ad}}}\right)^{-q} f(t_i)
\end{eqnarray}
where
\begin{eqnarray}
t_{min} = \left(\frac{\g}{\g_2}\right)^{1/k_{ad}} t \ .
\end{eqnarray}
The integral can be performed analytically.  If $t<Ct_s$,
\begin{eqnarray}
N(\g;t) = \frac{Kt^3\g^{-q}}{t_s^2(qk_{ad}+2)}
\left[ 1 - \left( \frac{\g }{\g_2}\right)^{qk_{ad}+2} \right]\ .
\end{eqnarray}
If $t>Ct_s$ and $\g t/\g_2<t_s$, 
\begin{eqnarray}
N(\g;t)  = K\g^{-q}
\nonumber \\ \times 
\Biggr[ \frac{t^3}{t_s^2(qk_{ad}+2)}
\left( \left( \frac{Ct_s}{t}\right)^{qk_{ad}+2} - 
\left(\frac{\g}{\g_2}\right)^{(qk_{ad}+2)/k_{ad}} \right) 
\nonumber \\ 
+ \frac{2t_s}{5(qk_{ad}-1)}
\left( 1 - \left(\frac{Ct_s}{t}\right)^{qk_{ad}-1}\right) \Biggr] \ .
\end{eqnarray}
If $t>t_s$ and $\g t/\g_2>t_s$, 
\begin{eqnarray}
N(\g;t) = \frac{2K \g^{-q} t_s}{5(qk_{ad}-1)} 
\left[ 1 - \left(\frac{\g}{\g_2}\right)^{(qk_{ad}-1)/k_{ad}}\right]\ .
\end{eqnarray}

\subsubsection{Solution with Radiative and Adiabatic Losses}

If adiabatic and cooling losses are important, then the cooling rate 
will have the form
\begin{eqnarray}
-\dot\g = \nu\ \g^2 + k_{ad}\frac{\g}{t}\ .
\end{eqnarray}
In this case, if $k_{ad}=1$, the
continuity Equation has the solution (see Appendix
\ref{appendix_both_soln})
\begin{eqnarray}
\g^2 N(\g;t) = \frac{K}{t} \int^{t}_{t_{min}} dt_i\ t_i\ \g_i^{2-q}\ f(t_i) 
\end{eqnarray}
where
\begin{eqnarray}
\g_i = \left[ t_i \left( (\g t)^{-1} - \nu\ln(t/t_i) \right) \right]^{-1}\ ,
\end{eqnarray}
\begin{eqnarray}
\label{tmin_eqn}
t_{min} = \left[ W\left( \frac{1}{\g_2 \nu t} e^{1/(\g \nu t)}\right)
  \g_2\ \nu\right]^{-1} \ ,
\end{eqnarray}
and $W(x)$ is the Lambert $W$ function, defined by
\begin{eqnarray}
x = W(x)e^{W(x)}\ 
\end{eqnarray}
\citep[e.g.,][]{corless96}.  The integral can be done analytically if
$q=2$.  In this case, for $t<t_s$,
\begin{eqnarray}
\g^2 N(\g;t) = \frac{K}{4\ t\ t_s^2} \left( t^4 - t_{min}^4\right)\ .
\end{eqnarray}
For $t>Ct_s$ and $t_{min}<Ct_s$, 
\begin{eqnarray}
\g^2 N(\g;t) = \frac{K}{t}\left[ \frac{1}{4t_s^2}( (Ct_s)^4 - t_{min}^4) + 
\frac{2t_s}{5} \left( t-Ct_s \right) \right]\ .
\end{eqnarray}
For $t>Ct_s$ and $t_{min}>Ct_s$, 
\begin{eqnarray}
\g^2 N(\g;t) = \frac{2K t_s}{5t} \left( t-t_{min} \right)\ .
\end{eqnarray}

\begin{figure}
\vspace{2.0mm}
\epsscale{1.0}
\plotone{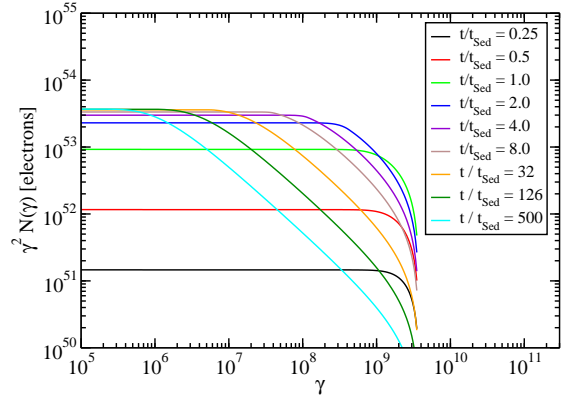}
\caption{Evolution of electron distribution for q=2 including
adiabatic and radiative cooling.  Parameters are given in 
Table \ref{param_table1}.}  
\label{electron3}
\vspace{2.0mm}
\end{figure}

The asymptotes of this solution can also be found.  We begin with the
case of $t<Ct_s$.  For $t \ll (\nu\g)^{-1}$ or $\g \ll (\nu t)^{-1}$,
the argument in the Lambert function in Equation (\ref{tmin_eqn}) goes
to infinity, so the Lambert function goes to infinity, and
$t_{min}\rightarrow 0$.  Then,
\begin{eqnarray}
\g^2 N(\g;t) \approx \frac{K t^3}{4 t_s^2} \propto\ t^3\ \g^0 \ .
\end{eqnarray}
For $t \gg (\nu\g)^{-1}$ or $\g \gg (\nu t)^{-1}$, a Taylor expansion 
of the exponential and Lambert function in Equation (\ref{tmin_eqn}) gives 
$t_{min} \rightarrow t - (\nu\g)^{-1}$.  Then 
\begin{eqnarray}
\g^2 N(\g;t) \approx \frac{K t^2}{16 t_s^2 \nu\g} \propto t^2\g^{-1}\ .
\end{eqnarray}
These results are quite similar to the asymptotes for the
radiative-only case (Section \ref{radiativelosses}).  The results for
$t>Ct_s$ are, however, somewhat different.  In this case, for $t \ll
(\nu\g)^{-1}$ or $\g \ll (\nu t)^{-1}$,
\begin{eqnarray}
\g^2 N(\g;t) \approx K t_s \left[ \frac{2}{5} - \frac{3t_s}{20t} \right]\ 
\approx \frac{2Kt_s}{5} \propto t^0 \g^0\ .
\end{eqnarray}
For $t \gg (\nu\g)^{-1}$ or $\g \gg (\nu t)^{-1}$, 
\begin{eqnarray}
\g^2 N(\g;t) \approx \frac{2Kt_s}{5\nu t \g}\ \propto t^{-1}\g^{-1}\ .
\end{eqnarray}
Here, the behavior for large $\g$ is essentially the same as the
radiative-only case.  However, the addition of adiabatic losses means
that the electrons at low $\g$ will approach a constant value, rather
than increasing logarithmically without bound.  This will remove the
inflection seen in the radiative-only case.  An example of this can be
seen in Figure \ref{electron3}, where it is clear the inflection in
the radiative losses-only solution (Section \ref{radiativelosses}) is
not seen.  Results are very similar for other values of $k_{ad}$,
particularly $k_{ad}=0.4$, as one would expect in the Sedov phase.

%\clearpage

\subsection{Spectral Energy Distribution}

Once the electron distribution has been determined, as above, the 
spectral energy distribution (SED) from the SNR can be calculated.  
For electrons in a randomly-oriented magnetic field $B$, the 
$\nu F_\nu$ synchrotron flux, $f_\e^{syn}$ is given by
\begin{eqnarray}
\label{fsy}
f_\e^{syn} = 
\frac{\sqrt{3} \e e^3 B}{4\pi h d^2} 
\int^\infty_1 d\g\ N_e(\g;t)\ R(x)\ ,
\end{eqnarray}
where $e=4.8\times10^{-10}$\ esu is the elementary charge, $\e =
h\nu/(m_e c^2)$ is the dimensionless observed photon energy, 
$h = 6.63\times10^{-27}\ \erg$-s is Planck's constant, $d$ is the 
distance to the SNR, 
\begin{equation}
x = \frac{4\pi m_e^2 c^3}{3eBh\g^2}\ ,
\end{equation}
\begin{eqnarray}
R(x) = \frac{x}{2}\int_0^\pi d\theta\ \sin\theta\ 
\int^\infty_{x/\sin\theta} dt\ K_{5/3}(t)\ 
\end{eqnarray}
\citep{crusius86}, and $K_{5/3}(t)$ is the modified Bessel function of
order 5/3.  Approximate expressions for $R(x)$ are given by
\citet{zirak07}; \citet*{finke08_SSC}; and \citet{joshi11}.

The electrons will also Compton-scatter external photon sources, such 
as the CMB or other intergalactic sources.  The flux from external 
Compton (EC) scattering a blackbody photon source with total energy 
density $u_{tot}$ and dimensionless temperature $\Theta=k_B T/(m_ec^2)$ is
given by
\begin{eqnarray}
\label{fccmb}
f_{\e}^{EC} = \frac{3m_e c\sT \e^2 }{16\pi d^2} \frac{15 u_{tot}}{(\pi\Theta)^4}
\\ \nonumber \times
\int_0^\infty d\e_*\ \frac{\e_*}{\exp(\e_*/\Theta)-1}
\\ \nonumber \times
\int_{\g_{min}}^{\g_{max}} d\g\ \g^{-2}\ N_e(\g;t)\ F_C(\e, \g, \e_*)
\end{eqnarray}
where $\sT = 6.65\times10^{-25}\ \cm^2$ is the Thomson cross section.
For the CMB at the present epoch, $\Theta=4.58\times10^{-10}$ and 
$u_{tot}=4.2\times10^{-13}\ \erg\ \cm^{-3}$ are the
dimensionless CMB temperature and energy density at the present epoch.  
The integration limits are given by 
\begin{eqnarray}
\g_{min} = \frac{1}{2}\e\left( 1 + \sqrt{1 + \frac{1}{\e_*\e}} \right)\ ,
\end{eqnarray}
\begin{eqnarray}
\g_{max} = \frac{\e_*\e}{\e_* - \e}H(\e_*-\e)\ ,
\end{eqnarray}
and the function
$$F_C(\e, \g, \e_*) =  \biggr[ 2q \ln q +(1+2q)(1-q) + 
%\nonumber \\
{1\over 2} {(\Gamma_e q)^2\over (1+\Gamma_e q)}(1-q) \biggr]$$
\begin{eqnarray}
\times H\;\left( q; {1\over 4\gamma^{2}}, 1 \right)\ ,
\end{eqnarray}
where
\begin{eqnarray}
q \equiv {\e/\g \over \Gamma_e
(1-\e/\g )}\;\;{\rm ,~ and}\; \; \Gamma_e = 4\e_*\g\; 
\end{eqnarray}
\citep{jones68,blumen70}.  The electrons can also Compton-scatter the
synchrotron photons produced by the same electron population (known as
synchrotron self-Compton or SSC), which is given by
$$
f_{\e}^{SSC} = \frac{9}{16} \frac{ \sT \e^{2}}
{\pi r^2 } $$
\begin{eqnarray}
 \times \int^\infty_0\ d\e_*\ 
\frac{f_{\e_*}^{syn}}{\e_*^{3}}\ \int^{\g_{max}}_{\g_{min}}\ d\g\ 
\frac{N_e(\g)}{\g^{2}} F_C(q,\Gamma)\ 
\end{eqnarray}
\citep[e.g.,][]{finke08_SSC}.  This mechanism is usually negligible
for large, diffuse remnants, but will play a roll in Section
\ref{multizone}.

The nonthermal electrons from the remnant will interact with the cold
ions (assumed to be protons) in the surrounding CSM to make
bremsstrahlung (or free-free radiation) with flux given by
$$f_\e^{ff} = \frac{n_{CSM} m_e c^3 \e^2}{4\pi d^2}$$
\begin{eqnarray}
\times 
\int_1^{\infty}\ d\g\ N_e(\g;t)\ \frac{d\sigma_{ff,eZ}}{d\e}(\e;\g)\ ,
\end{eqnarray}
where the bremsstrahlung cross section is written as 
$$\frac{d\sigma_{ff,eZ}}{d\e}(\e;\g) = $$
\begin{eqnarray}
\frac{4Z^2\alpha_f r_e^2}{\e}
\left( 1 + y - \frac{2y}{3}\right)
\left[ \ln\left( \frac{2\g^2 y}{\e}\right) - \frac{1}{2} \right]
\end{eqnarray}
\citep{blumen70}, $Z$ is the effective charge of the cold ions,
$\alpha_f \approx (137)^{-1}$ is the fine structure constant, $r_e =
2.82\times10^{-13}\ \cm$ is the classical electron radius, and $y =
1-\e/\g$.

\begin{figure}
\vspace{2.0mm}
\epsscale{1.0}
\plotone{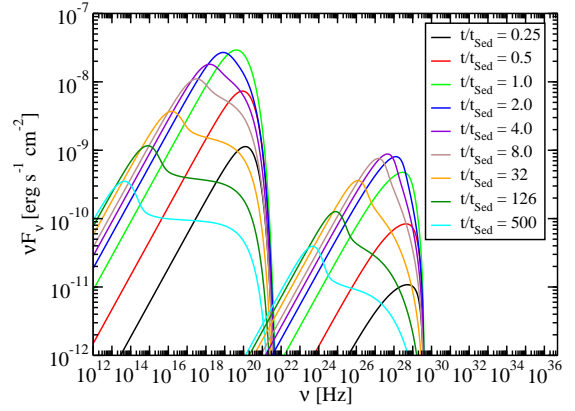}
\caption{Evolution of synchrotron and Compton-scattered CMB flux for
the electron distribution in Figure \ref{electron1}.
}
\label{SED1}
\vspace{2.0mm}
\end{figure}

\begin{figure}
\vspace{2.0mm}
\epsscale{1.0}
\plotone{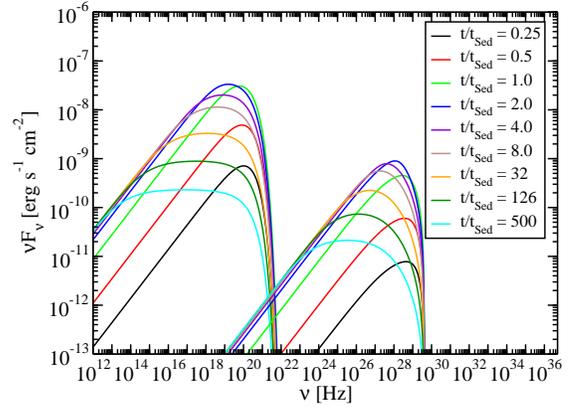}
\caption{Evolution of synchrotron and Compton-scattered CMB flux for
the electron distribution in Figure \ref{electron3}.
}
\label{SED3}
\vspace{2.0mm}
\end{figure}

The synchrotron and Compton emission can be seen in Figs.\ \ref{SED1}
and \ref{SED3}, corresponding to the respective electron distributions
seen in Figs.\ \ref{electron1} and \ref{electron3}.  These SEDs assume
emission from an SNR at $d=1\ \kpc$.  In the Sedov phase, the feature
resulting from the different evolution above and below the cooling
break is clearly seen for the case when adiabatic losses are neglected
(Figure \ref{SED1}), but this feature is not as pronounced when
adiabatic losses are taken into account (Figure \ref{SED3}).

%\clearpage

%\begin{figure}
%\epsscale{1.0}
%\plotone{flux_q3}
%\caption{Evolution of synchrotron and Compton-scattered CMB flux for q=3.  
%}
%\label{SED2}
%\end{figure}
%\clearpage

%\clearpage

\section{Application to the Remnant \rxj}
\label{rxj_section}

We apply our results to SNR \rxj\ (G~347.3$-$0.5), which is thought to
be the remnant of a ``guest star'' observed by Chinese astronomers in
393 CE \citep{wang97}.  This fixes it age at $\cong 1620$ yr, so SNR
\rxj\ is likely to be well into the Sedov phase, although
\citet{fukui03} argue instead that it is still in the free expansion
stage, for CSM densities $n_{CSM}< 0.01$ cm$^{-3}$ at $\approx 1$ kpc
distance.  \citet{slane99} associated the source with a nearby
molecular cloud, giving a distance to the source of $\sim 6\ \kpc$.
However, newer CO observations found molecular gas at $\sim 1\ \kpc$,
\citep{fukui03,moriguchi05}.  Absorbing column densities from X-ray
observations strengthen the $\sim 1\ \kpc$ distance estimate
\citep{koo04,cassam04}, making it the most likely one.  The detection
of an X-ray point source, thought to be a left-over neutron star,
implies the remnant is the result of a core-collapse supernova
\citep{lazendic03}.

The X-ray spectrum of \rxj\ appears completely dominated by nonthermal
emission \citep[e.g.,][]{tanaka08}.  The lack of thermal X-ray lines
is taken as evidence that $n_{CSM}\la 0.2\ \cm^{-3}$
\citep{slane99,ellison10,abdo11_rxj}.  Without a dense target for
cosmic ray protons, $\pi^0$ decay is probably not a significant
contributor to the $\g$-ray spectrum of this remnant
\citep{ellison10}, and electron bremsstrahlung is weak in comparison
with the Compton $\gamma$-ray emission.  But note the different energy
ranges of electrons that radiate into the $\gamma$-ray band: electrons
with Lorentz factor $\gamma \sim 10^6$ ($\sim 10^8$) scatter CMB
photons to GeV (10 TeV) ranges, and electrons with $\gamma\sim 10^3$
($\sim 10^6$) make GeV (TeV) bremsstrahlung.

HESS observations show that the X-ray image of \rxj\ and the VHE
$\g$-rays are spatially well-correlated
\citep{aharonian04_rxj,aharonian06_rxj}.  \citet{uchiy07} reported
X-ray variability on a year timescale in a few small (arcsecond scale)
hotspots of \rxj.  If this reflects radiative variability of the
nonthermal electrons, large magnetic fields are needed ($B\sim 1$\
mG), and the implied small number of electrons in such a strong
magnetic field could not produce the TeV emission. But the appearance
of thin radio-emitting rims in some SNRs could mean that the emission
zone is compact and variability is due to shell expansion or
compression as it encounters dilute or dense CSM
\citep{reynolds10,reynolds11}.  Furthermore, the existence of knots
which do not seem to be variable on such short timescales indicates
knots may exist with significantly lower magnetic fields.  Based on
the HESS observations, \citet{aharonian06_rxj} concluded that a
leptonic model was unlikely to fit the broadband SED, and a hadronic
origin was favored for the $\gamma$ rays from \rxj.  The variable
X-ray filaments cannot, however, explain global TeV emission
\citep{butt08}.

With the arrival of the first epoch \fermi\ data from \rxj\
\citep{abdo11_rxj}, different models can be tested better.  This seems
to make it a good time to revisit leptonic models, which are natural
for \rxj\ given the close spatial correlation between the X-ray and
TeV $\gamma$ rays.

\subsection{Single Zone Model Fit}
\label{singlezone}

\begin{figure}
\vspace{2.0mm}
\epsscale{1.0}
\plotone{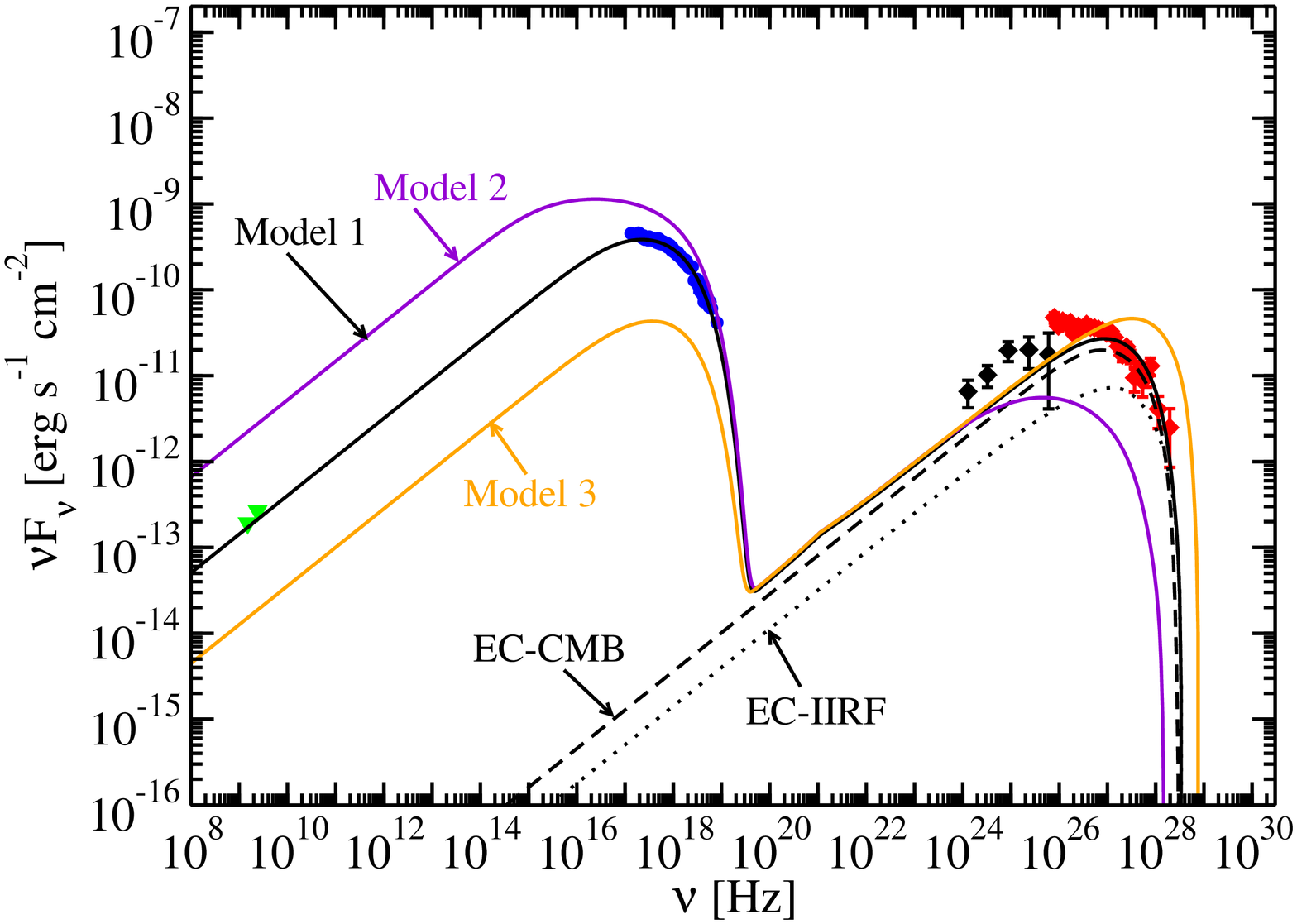}
\caption{The integrated broadband SED of \rxj\ with data from ATCA
\citep[green triangles;][]{aharonian06_rxj}, {\em Suzaku} \citep[blue
circles;][]{tanaka08}, LAT \citep[black diamonds;][]{abdo11_rxj}, and
HESS \citep[red diamonds;][]{aharonian06_rxj,aharonian07_rxj}.  The
solid curves show the model fits for different magnetic fields, as
labeled.  The dashed and dotted curves show the Compton-scattered CMB
and IIRF components, respectively, for Model 1.
}
\label{rxj1713_B}
\vspace{2.mm}
\end{figure}

The integrated broadband SED of \rxj\ is shown in Fig.\
\ref{rxj1713_B}.  \citet{porter06} have shown that the interstellar
infrared radiation field (IIRF) can be a significant photon source for
Compton scattering.  It is strongly dependent on the position in the
Galaxy, and close to the Galactic center their model
\citep{moskalenko06} gives the IIRF energy density greater than that
of the CMB.  At a Galactic longitude of $\ell=347.3\arcdeg$ and a
Galactic latitude of $b=-0.5\arcdeg$, this remnant is nearly along
the line of sight of the Galactic center.  This makes the intensity of
the IIRF strongly dependent on its distance from the Earth (and
therefore the Galactic center).  The consensus, based on associations
of molecular clouds and absorption of X-rays (as discussed above, in
Section \ref{rxj_section}) seems to be that the \rxj is at $d=1\ \kpc$
from Earth.  Therefore in our modeling we use this distance, and an
IIRF intensity consistent with this distance (or $7.5\ \kpc$ from the
Galactic center) from \citet{porter06}, modeled as a blackbody with
temperature $T=30\ \Kelvin$ and total energy density
$u=4.8\times10^{-13}\ \erg\ \cm^{-3} = 0.30\ \eV\ \cm^{-3}$.  Note
that \citet{li11} modeled the source using the IIRF at a level near
what one would expect \citep{porter06} if \rxj\ was $d=6\ \kpc$ from
the Earth.  

The models in Figure \ref{rxj1713_B} include adiabatic and radiative
losses from both the CMB and IIRF, and apply to the integrated
emission over the entire remnant.  In our modeling we have added
this radiation field as an additional term in equation (\ref{nu_cool})
to take it into account in the evolution of the SNR.  We take the age
of the remnant to be $t=1620\ \yr$.  A best fit to the SED, given
reasonable parameter constraints on age, ICM density, initial mass,
and blast energy, is shown in the figure as the black curve.  This
model includes emission from synchrotron, Compton-scattering of CMB
and IIRF photons, and bremsstrahlung, which can be seen as a small
bump at $\sim10^{21}\ \Hz$.  It provides a good fit to the radio and
X-ray data, but does not adequately reproduce the LAT and lower-energy
HESS measurements.

The model parameters are shown in Table \ref{param_table2}.  The
parameters which are constrained by the SED fit are $B$, $q$, $\g_2$,
and $\eta_e$.  We also display in this Figure models with 5 times
larger and small $B$.  As the magnetic field increases, the overall
synchrotron flux increases, and the cooling electron Lorentz factor
$(\nu t)^{-1}$ decreases.  The magnetic field will only affect the
Compton-scattered emission through its effects on the electron
distribution; thus the overall Compton-scattered flux will not
increase, and indeed will decrease above the the cooling break, which
is lower for higher $B$.

%\clearpage
\begin{deluxetable*}{lcccccc}
%\begin{deluxetable}{lcccccc}
\tabletypesize{\scriptsize}
\tablecaption{
\rxj\ Model Parameters
}
\tablewidth{0pt}
\tablehead{
\colhead{Parameter} &
\colhead{Symbol} &
\colhead{Model 1 (baseline) } &
\colhead{Model 2 } &
\colhead{Model 3} &
\colhead{Model 4} &
\colhead{Model 5} 
}
\startdata
Blast Energy [erg]& $E$ & $1.6\times10^{51}$ & $1.6\times10^{51}$ & $1.6\times10^{51}$ & $1.6\times10^{51}$ & $1.6\times10^{51}$ \\
Initial Mass [$M_\odot$] & $M_0$  & 1.6 & 1.6 & 1.6 & 6.4 & 0.4 \\
Initial Velocity [$\cm\ \s^{-1}$] & $v_0$  & $1.0\times10^9$ & $1.0\times10^9$ & $1.0\times10^9$ & $5\times10^8$ & $2.0\times10^9$ \\
ICM density [$\cm^{-3}$] & $n_{ICM}$  & 0.2 & 0.2 & 0.2 & 0.2 & 0.2 \\ 
Sedov time [yr] & $t_{s}$ & 420 & 420 & 420 & 1300 & 132 \\
\hline
Magnetic field [$\mu$G] & $B$ & 12 & 60 & 2.4 & 12 & 12 \\
Cooling Constant [$\s^{-1}$] & $\nu$ & $2.2\times10^{-19}$ & $4.7\times10^{-18}$ & $3.7\times10^{-20}$ & $2.2\times10^{-19}$ & $2.2\times10^{-19}$\\
Cooling electron Lorentz factor & $(\nu t)^{-1}$ & $9.1\times10^7$ & $4.2\times10^6$ & $5.4\times10^8$ & $9.1\times10^7$& $9.1\times10^7$ \\
\hline
%Maximum Acceleration efficiency & $\phi$ & $5.3\times10^{-5}$ & $5.3\times10^{-5}$ & $5.3\times10^{-5}$ \\ 
Low energy electron cutoff & $\g_1$ & 10 & 10 & 10 & 10 & 10 \\ 
High energy electron cutoff & $\g_2$ & $3.1\times10^8$ & $3.1\times10^8$ & $3.1\times10^8$ & $3.1\times10^8$ & $3.1\times10^8$ \\
Injection spectral index & $q$ & $2.1$ & $2.1$ & $2.1$ & $2.1$ & $2.1$ \\
Electron acceleration efficiency & $\eta_e$ & $5.0\times10^{-5}$ & $5.0\times10^{-5}$ & $5.0\times10^{-5}$ & $5.0\times10^{-5}$ & $5.0\times10^{-5}$ \\
\enddata
\label{param_table2}
\vspace{2.0mm}
\end{deluxetable*}
%\end{deluxetable}
%\clearpage

\begin{figure}
\vspace{4.0mm}
\epsscale{1.0}
\plotone{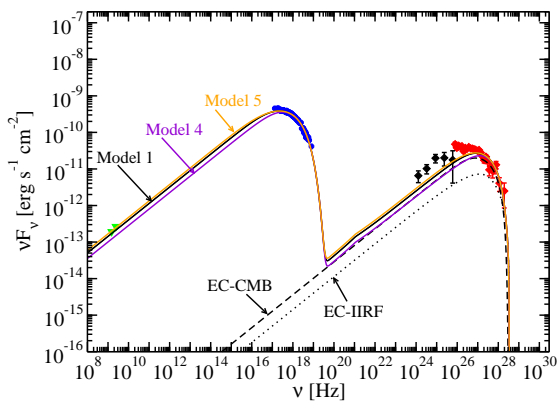}
\caption{ Similar to Fig.\ \ref{rxj1713_B}, only $v_0$ is varied instead 
of $B$.  
}
\label{rxj1713_v}
\vspace{2.mm}
\end{figure}

In Fig.\ \ref{rxj1713_v} we explore variations in $v_0$, which also
correspond to variations in $M_0$, assuming $E$ is held constant, as
given in Equation (\ref{energy}).  The Sedov radius $r_s \propto
v_0^{-2/3}$ as given in Equation (\ref{rsedov}), and thus $t_s \propto
v_0^{-5/3}$ (Equation [\ref{tsedov}]) and $K\propto v_0^{5/3}$
(Equation [\ref{Kconst}]).  Keeping this in mind, $t>t_s$ so that
below the cooling break, $N_e(\g;t) \propto 1/3+(5/2)\ln(v_0^{5/3})$
from Equation (\ref{g2N_1}) and above the break, $N_e(\g;t) \propto
v_0^0$ (Equation [\ref{g2N_2}]).  This is reflected in the SED, as
seen in Fig.\ \ref{rxj1713_v}, where below the break the emission
increases with $v_0$, while above the break, the emission is
independent of $v_0$.  Also note that it is unlikely for the initial
ejecta mass to be as low as found in Model 4, demonstrating the
limited usefulness in varying $v_0$ to obtain a good fit to \rxj.

%\clearpage

%\clearpage

\subsection{Multi-Zone model}
\label{multizone}

As discussed above in Section \ref{rxj_section}, the discovery of
variable X-ray filaments (or knots) within the SNR structure by
\citet{uchiy07} indicates that the single-zone fit is inadequate to
explain the overall SED.  However, the filaments themselves could
contribute a significant amount to the $\g$-ray emission from the
source.  

\begin{figure}
\vspace{2.0mm}
\epsscale{1.0}
\plotone{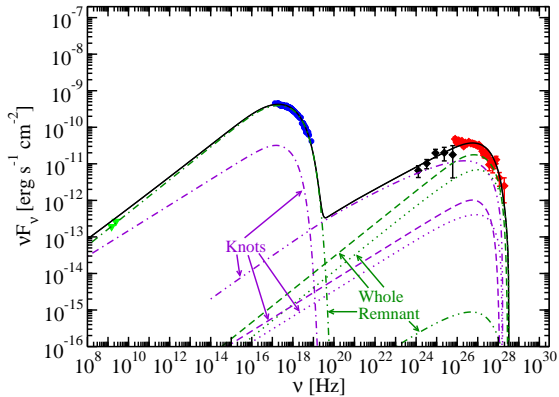}
\caption{ Multi-zone model fit to \rxj.  Curves show the total
emission from the knots and overall shock combined (black solid curve)
as well as synchrotron emission (dot-dashed curves),
Compton-scattered CMB (dashed curves), Compton-scattered IIRF (dotted
curves), and SSC (double dot-dashed curves) from the overall shock and
knots.  }
\label{rxj1713_multi}
\vspace{2.mm}
\end{figure}
%\clearpage

Smaller knots emitting synchrotron, SSC, and Comptonized CMB and IIRF
radiation were added to Model 1, as seen in Figure
\ref{rxj1713_multi}.  The much smaller volume of these knots results
in large synchrotron energy densities in the knots, with strong SSC
emission at GeV energies.  This fit has the number of zones taken to
be $N_{knots}=100$, with each zone having $B_{knots}=16\ \mu$G, radii
$R_{knot}=1\ $mpc, and an electron distribution that spans from
$\g_{knot,1}=10$ to $\g_{knot,2}=1.4\times10^8$ with a break at
$\g_{knot,brk}=4.7\times10^7$ with $N_{e,knot}(\g)\propto\g^{-2.3}$ for
$\g<\g_{knot,brk}$ and $N_{e,knot}(\g)\propto\g^{-3.3}$ for
$\g>\g_{knot,brk}$.  As can be seen in Figure \ref{rxj1713_multi},
this reproduces the SED well, and makes interesting predictions.

The synchrotron component is dominated by the large first zone that
effectively represents the entire remnant, which also makes the bulk
of the TeV radiation. Emission $\ga 1\ \TeV$ is dominated by the
Compton-scattered CMB of the remnant as a whole, while in the range in
the joint LAT/HESS window from $\la 1\ \TeV$ the $\g$ rays arise from
the SSC component in the knots.  The angular resolution of the LAT is
generally worse than $0.1\arcdeg$.  At a distance of 1 kpc, the 1
mpc knots will have an angular radius of $0.2\arcsec$ and thus cannot
be resolved with the LAT.  CTA will have an angular resolution of
$\sim 1\arcmin$\ \citep{cta10} and will not be able to distinguish the
variable and non-variable X-ray knots seen by \citet{uchiy07} either,
even if they radiate in $\g$ rays.  However, if the low and high
energy $\g$-rays come from different components, maps of \rxj\ made
with CTA may be different at lower ($\la 1\ \TeV$) and higher ($\ga 1\
\TeV$) energies, with the higher energy maps being more in agreement
with X-ray ones.  This may allow this multi-zone model to be tested.

The knots contribute $\sim10$\% to the X-ray emission of the remnant,
consistent with observations from \citet{uchiy03}.  They are also much
lower than the values inferred from variability by \citet{uchiy07}.
However, there seem to be many knots which are not variable, which
could reflect a lower magnetic field.

\section{Discussion and Conclusions}
\label{conclusion}

The SNR \rxj\ occupies an important place in $\gamma$-ray studies of
supernova remnants.  TeV emission from the SNR \rxj\ was first
detected with the CANGAROO experiment \citep{muraishi00}.  Based on
further CANGAROO observations, \citet{enomoto02} claimed that a
standard leptonic synchrotron/EC-CMB model did not fit these data,
including the EGRET upper limit. \citet{reimer02} argued that EGRET
upper limits rule out a hadronic origin, but diffusion of high-energy
particles upstream of the shock can harden nuclear emission
\citep{malkov06}. \citet{aharonian04_rxj} produced the first resolved
$\g$-ray image of an SNR by HESS.  Further HESS observations found
that the X-ray and VHE $\g$-rays were spatially well-correlated
\citep{aharonian06_rxj}.  \citet{porter06} found, however, that
Compton-scattered Galactic background photons, in addition to CMB
photons, could help to explain the \rxj\ VHE emission in leptonic
models.  Still further HESS observations detected the remnant out to
$\sim 100~ \TeV$ \citep{aharonian07_rxj}.  \citet{li11} provide a good
fit to the full SED including the LAT spectrum with a model similar to
\citet{porter06}, including Compton-scattering of interstellar
infrared photons.  As discussed above in Section \ref{singlezone},
they assumed the source was at a distance of 6 kpc from us, closer to
the Galactic center where the IIRF is much more intense.  However, we
think the molecular cloud and X-ray absorption evidence points to \rxj\ 
most likely being at $d=1\ \kpc$.  This emphasizes the crucial
importance of an accurate distance measurement to SNR modeling.

New data from \fermi\ \citep{abdo11_rxj}, in addition to
multiwavelength measurements at radio, X-ray, and TeV energies, reveal
the bolometric SED of SNR \rxj\ with unprecedented detail.  The joint
\fermi-LAT/HESS data favor models
\citep{porter06,berezhko06,ellison10,zirak10} where the $\g$ rays have
a leptonic rather than a hadronic origin.  This conclusion follows
rather forcefully if the injection spectral index of the
particles---protons or electrons---is softer than $q=2$, as expected
in the linear first-order Fermi acceleration theory.  Nonlinear
effects may modify the injection index \citep[e.g.,][]{blasi05}.
Indeed, \citet{yamazaki09} point out that nonlinear effects could
harden the emission in the LAT energy range, making it nearly
impossible to distinguish between leptonic and hadronic origins.
\citet{inoue11} find that hard $\g$-ray spectra can be generated from
$\pi^0$ decay if the CSM has inhomogeneities and is ``clumpy''.

These results also test the conclusions of \citet{fukui11} based on a
comparison of the TeV and CO and \ion{H}{1} morphology.  They
interpreted the good correlation between the two bands as being a
strong signature for a hadronic origin of the $\g$ rays, since cosmic
ray protons would react with the molecular cloud hadrons.  That
interpretation is not unique to hadrons, however, as shocks in
molecular clouds would enhance electron acceleration and leptonic
$\g$-ray emission. Observations with HAWC and CTA, and longer
exposures with the {\em Fermi}-LAT, will show the fraction of
radiation from clumps at different energies, and will help clarify the
issue.  Here we consider broadband spectral modeling by following
electron injection and evolution.

The complicated CSM distribution in any realistic SNR environment is
quite different from the assumption of a homogeneous medium, but
within this approximation, we reconsidered particle injection, and
found that the assumption that the injection power is proportional to
the rate at which kinetic energy is swept downstream of an adiabatic
blast wave yields interesting structure in the particle injection
distribution that is cooling. The addition of adiabatic losses
significantly smooths these effects, but in either case, fitting of
the \rxj\ data with a single-zone synchrotron/Compton-scattered model
did not give a perfect fit.

The addition of knots, as in the two-zone models of
\citet{atoyan00_gammaray,atoyan00_twozone} applied to Cas A, introduce
interesting effects on electrons escaping downstream into a region of
different magnetic field.  In a two-zone model, particles may be
accelerated in smaller knots, and diffuse into a larger zone.  The
model of \citet{atoyan00_gammaray,atoyan00_twozone} is justified by
the small knots seen in the 6.3 cm Very Large Array (VLA) image of the
remnant Cas~A.  The VLA or another high angular resolution radio
telescope has not yet observed \rxj, although we justify this
complication from knots observed from this source in X-rays
\citep{uchiy07}.  \citet{bykov08} showed that the structure and
variability in these X-ray images can be reproduced with a steady
electron population in a random magnetic field.
Similar variable (on $\sim 4$\ year timescales)
X-ray knots have been found in Cas~A \citep{uchiy08}, also implying
large fields ($B\sim 1$\ mG) similar to \rxj, assuming the variability
is due to radiative cooling.  Cas~A is the result of a IIb supernova
\citep{krause08}, and \rxj\ is probably the result of a core-collapse
supernova \citep{lazendic03}, so they seem to be of similar type,
although Cas~A is much younger than \rxj, $\sim 300$\ years
\citep{fesen06} versus $\sim 1600$\ years.  Note that
\citet{yamazaki09} have also applied two zone leptonic and hadronic
models to \rxj, finding that one-zone models could not explain the VHE
$\g$-ray spectrum and LAT upper limits available at the time.

\citet{katz08} use radio observations of SNRs in nearby galaxies to
put a lower limit on the ratio of accelerated electrons to protons.
Assuming this ratio is approximately the same for all SNRs, they find
that a hadronic explanation for the $\g$-ray emission from \rxj\ is
unlikely.  \citet{yuan11} have fit the broadband SED of \rxj\ with
three models, consisting of leptonic, hadronic, and hybrid
leptonic/hadronic emission.  They found their hadronic model provided
the best fit, but it also had the greatest number of
poorly-constrained free parameters.  Because of this, along with
requiring an unrealistically large amount of energy put into
nonthermal protons, they concluded that they could not distinguish
between their three scenarios.

Based on spectral modeling of the broadband SED of the Tycho SNR,
particularly the shape of the \fermi-LAT and VERITAS spectra, it has
been suggested that only hadronic emission, and not leptonic emission,
can be the source of $\g$ rays from this object
\citep{morlino11,giordano11}.  However, this has been called into
question by a two-zone model \citep{atoyan11}. Final conclusions
regarding cosmic-ray proton/ion acceleration in Tycho rest on the
spectral shape below $\approx 400$ MeV. A two-zone leptonic model for
\rxj, as we have seen here, avoids any need for cosmic-ray proton
acceleration.  Note that this does not preclude cosmic-ray proton
acceleration either, simply that protons do not contribute
significantly to the emitted electromagnetic radiation.

In conclusion, we have described a simple model for the time evolution
of SNR emission.  This includes an assumption that particle
acceleration efficiency is proportional to the power swept into the
expanding blast wave.  Effects of radiative and adiabatic cooling on
the evolving particle distribution, and emission from synchrotron,
bremsstrahlung, and Compton-scattering processes were taken into
account.  In doing this, we have made a number of simplifying
assumptions.  We have assumed the CSM density is constant, which may
be less likely for remnants of core-collapse rather than Type Ia
supernovae.  We have assumed the magnetic field strength and power-law
injection index do not vary with time, and have used a simple Sedov
solution neglecting reverse shocks.  Hadronic emission processes were
neglected in this study.  We have applied our evolution model to \rxj,
and showed that a single-zone model cannot reproduce its SED if it 
is at a distance $d=1\ \kpc$.  The
addition of a second zone consisting of compact knots gives an
acceptable fit and makes interesting radio and $\gamma$-ray
predictions that should be testable in the near future.

\acknowledgements 

We are grateful to T.\ Tanaka for sending us the {\em Suzaku} spectral
data for \rxj, S.\ Funk for the {\em Fermi}-LAT spectral data on the
same source, and A.\ Atoyan for useful conversations regarding
energetics and multi-zone modeling in SNRs.  We would also like to
thank S.\ Reynolds and R.\ Yamazaki for useful correspondence
regarding their work on SNRs, F.\ Acero for pointing out an
important error in a previous version of this manuscript, and the
anonymous referee for a helpful and constructive report.  This work
is supported by the Office of Naval Research.

\appendix

\section{Solution to the Continuity Equation for Radiative Cooling Only}
\label{appendix_radonly_soln}

We wish to solve the continuity equation, Equation (\ref{cont_eqn}), for 
$t_{esc}\rightarrow\infty$ and $\dot\g=-\nu\g^2$:
\begin{eqnarray}
\frac{\partial N}{\partial t} - 
\nu\ \frac{\partial}{\partial\g}\left[ \g^2\ N(\g;t)\right] 
 = Q(\g,t)\ .
\end{eqnarray}
This has the solution
\begin{eqnarray}
\label{Ngt_soln}
N(\g;t) = \int_0^{t} dt_i\ \int_{\g_1}^{\g_2}\ d\g_i\ G(\g_i,t_i,\g;t)\ 
Q(\g_i,t_i)
\end{eqnarray}
where $\g_1$ and $\g_2$ are the respective lower and upper limits on
the injected electrons' Lorentz factors, $G(\g_i,t_i,\g;t)$ is the
Green's function which satisfies the equation
\begin{eqnarray}
\label{green_cont_eqn}
\frac{\partial G}{\partial t} - 
\nu\ \frac{\partial}{\partial\g}\left[ \g^2\ G\right] 
 = \delta(\g-\g_i)\ \delta(t-t_i)\ 
\end{eqnarray}
and $\delta(x)$ is the standard Dirac delta function.  A 
single particle injected with Lorentz factor $\g_i$ 
at time $t_i$ will, at some later time $t$, have a 
Lorentz factor $\g(t)$ which satisfies the equation
\begin{eqnarray}
\label{dgdt_eqn}
\frac{d\g}{dt} = - \nu\ \g^2\ .
\end{eqnarray}
Equation (\ref{dgdt_eqn}) can be solved, 
\begin{eqnarray}
\label{gamma_rad}
\g(t) = \frac{1}{\g_i^{-1} + \nu(t-t_i)}\ ,
\end{eqnarray}
and thus the solution to Equation (\ref{green_cont_eqn}) is
\begin{eqnarray}
G(\g_i,t_i,\g;t) = \delta\{\g-[\g_i^{-1} + \nu(t-t_i)]^{-1}\}\ ,
\end{eqnarray}
which can be rewritten as
\begin{eqnarray}
G(\g_i,t_i,\g;t) = \frac{\g_i^2}{\g^2}\ 
\delta\{\g_i-[\g^{-1} - \nu(t-t_i)]^{-1}\}\ .
\end{eqnarray}
Inserting this into Equation (\ref{Ngt_soln}), one can perform the 
integral over $\g_i$ to get
\begin{eqnarray}
N(\g;t) = \frac{1}{\g^2}\ \int_{t_{min}}^t\ dt_i 
\left[ \g^{-1} - \nu(t-t_i) \right]^{-2}\ 
Q\left[ \left\{\g^{-1} - \nu(t-t_i)\right\}^{-1}, t_i \right]\ ,
\end{eqnarray}
where
\begin{eqnarray}
t_{min} = \max[0, t - \nu^{-1}(\g^{-1} - \g_2^{-1})]\ .
\end{eqnarray}
The lower limit $t_{min}$ comes about because particles are 
injected only with $\g_i<\g_2$.

\section{Solution to the Continuity Equation for Radiative and Adiabatic Cooling}
\label{appendix_both_soln}

We now wish to solve the continuity equation, Equation (\ref{cont_eqn}) 
for radiative and adiabatic cooling, i.e., 
\begin{eqnarray}
\label{dgdt_both}
-\frac{d\g}{dt} = \nu\g^2 + k_{ad} \frac{\g}{t}\ .
\end{eqnarray}
We can follow the same procedure as in Appendix \ref{appendix_radonly_soln}.  
Equation (\ref{dgdt_both}) can be solved for $\g(t)$ to give 
\begin{equation}
\label{gamma_both}
\g(t) = 
\left\{ t^{k_{ad}}\left[ (\g_i t_i^{k_{ad}})^{-1} + 
\nu\ T(t,t_i)
\right]\right\}^{-1} 
\end{equation}
\citep{gupta06} where
\begin{equation}
T(t,t_i) = \left\{ \begin{array}{ll}
(t^{1-k_{ad}} - t_i^{1-k_{ad}})/(1-k_{ad})  & k_{ad} \ne 1 \\
\ln(t/t_i) & k_{ad}=1 
\end{array}\ .
\right.
\end{equation}
In this case Equation
(\ref{gamma_both}) implies the Green's function which satisfies
Equation (\ref{green_cont_eqn}) is
\begin{eqnarray}
G(\g_i,t_i,\g;t) = \frac{1}{ \g^2\ t^{k_{ad}}\ t_i^{k_{ad}} \left[ (\g t)^{-1} - 
\nu\ T(t,t_i) \right]^2 }\ 
\delta\left[ \g_i - \frac{1}{ t_i^{k_{ad}}\left[ (\g t^{k_{ad}})^{-1} - 
\nu\ T(t,t_i)\right] } \right]\ .
\end{eqnarray}
Substituting this into Equation (\ref{Ngt_soln}) and performing the integral 
over $\g_i$ with the help of the Dirac $\delta$-function gives
\begin{eqnarray}
N(\g;t) = \frac{K}{t^{k_{ad}}\ \g^2}\ \int_{t_{min}}^{t}\ dt_i\ t_i^{k_{ad}}\ f(t_i)\ 
\left\{\frac{1}{ t_i\left[ (\g t)^{-1} - 
\nu\ T(t,t_i)\right] }\right\}^{2-q}\ .
\end{eqnarray}
The lower limit $t_{min}$ can be found from the constraint that
\begin{eqnarray}
\label{inequal_lowlimit}
\frac{1}{ t_i\left[ (\g t)^{-1} - 
\nu\ T(t,t_{i})\right]} < \g_2\ .
\end{eqnarray}
For $k_{ad}=1$, when solved for $t_i$ this constraint gives
\begin{eqnarray}
t_{min} = \left[ W\left( \frac{1}{\g_2 \nu t} e^{1/(\g \nu t)}\right)
  \g_2\ \nu\right]^{-1} \ ,
\end{eqnarray}
where $W(x)$ is the Lambert W function \citep{corless96}.  For general 
values of $k_{ad}$, Equation (\ref{inequal_lowlimit}) does not 
have a simple analytic solution, and it is solved numerically for $t_{min}$.

If $k_{ad}=0$, i.e., there are no adiabatic losses, then Equation
(\ref{gamma_both}) will reduce to Equation (\ref{gamma_rad}),
leading to the radiative losses-only solution.  On the other hand, if
radiative losses are negligible, i.e. $\nu\rightarrow 0$, then
Equation (\ref{gamma_both}) will reduce to $\g(t)= \g_i
(t_i/t)^{k_{ad}}$ which leads to the solution in Section
\ref{adiabaticlosses}.

%!****************************************************

%\begin{thebibliography}{}

\bibliographystyle{apj}
\bibliography{SNR_reference,RXJ1713_reference,blazar_ref,IC443_reference,CasA_reference}

\begin{thebibliography}{87}
\expandafter\ifx\csname natexlab\endcsname\relax\def\natexlab#1{#1}\fi

\bibitem[{{Abdo} {et~al.}(2009)}]{abdo09_w51c}
{Abdo}, A.~A., {et~al.} 2009, \apjl, 706, L1

\bibitem[{{Abdo} {et~al.}(2010{\natexlab{a}})}]{abdo10_1fgl}
---. 2010{\natexlab{a}}, \apjs, 188, 405

\bibitem[{{Abdo} {et~al.}(2010{\natexlab{b}})}]{abdo10_w44}
---. 2010{\natexlab{b}}, Science, 327, 1103

\bibitem[{{Abdo} {et~al.}(2010{\natexlab{c}})}]{abdo10_ic443}
---. 2010{\natexlab{c}}, \apj, 712, 459

\bibitem[{{Abdo} {et~al.}(2011{\natexlab{a}})}]{abdo11_2fgl}
---. 2011{\natexlab{a}}, \apj, submitted, arXiv:1108.1435

\bibitem[{{Abdo} {et~al.}(2011{\natexlab{b}})}]{abdo11_rxj}
---. 2011{\natexlab{b}}, \apj, 734, 28

\bibitem[{{Aharonian} {et~al.}(2006)}]{aharonian06_rxj}
{Aharonian}, F., {et~al.} 2006, \aap, 449, 223

\bibitem[{{Aharonian} {et~al.}(2007)}]{aharonian07_rxj}
---. 2007, \aap, 464, 235

\bibitem[{{Aharonian} \& {Atoyan}(1996)}]{aharonian96}
{Aharonian}, F.~A., \& {Atoyan}, A.~M. 1996, \aap, 309, 917

\bibitem[{{Aharonian} {et~al.}(2004)}]{aharonian04_rxj}
{Aharonian}, F.~A., {et~al.} 2004, \nat, 432, 75

\bibitem[{{Atoyan} {et~al.}(2000{\natexlab{a}}){Atoyan}, {Aharonian}, {Tuffs},
  \& {V{\"o}lk}}]{atoyan00_gammaray}
{Atoyan}, A.~M., {Aharonian}, F.~A., {Tuffs}, R.~J., \& {V{\"o}lk}, H.~J.
  2000{\natexlab{a}}, \aap, 355, 211

\bibitem[{{Atoyan} \& {Dermer}(2011)}]{atoyan11}
{Atoyan}, A.~M., \& {Dermer}, C.~D. 2011, submitted

\bibitem[{{Atoyan} {et~al.}(2000{\natexlab{b}}){Atoyan}, {Tuffs}, {Aharonian},
  \& {V{\"o}lk}}]{atoyan00_twozone}
{Atoyan}, A.~M., {Tuffs}, R.~J., {Aharonian}, F.~A., \& {V{\"o}lk}, H.~J.
  2000{\natexlab{b}}, \aap, 354, 915

\bibitem[{{Badenes} {et~al.}(2006){Badenes}, {Borkowski}, {Hughes}, {Hwang}, \&
  {Bravo}}]{badenes06}
{Badenes}, C., {Borkowski}, K.~J., {Hughes}, J.~P., {Hwang}, U., \& {Bravo}, E.
  2006, \apj, 645, 1373

\bibitem[{{Baring} {et~al.}(1999){Baring}, {Ellison}, {Reynolds}, {Grenier}, \&
  {Goret}}]{baring99}
{Baring}, M.~G., {Ellison}, D.~C., {Reynolds}, S.~P., {Grenier}, I.~A., \&
  {Goret}, P. 1999, \apj, 513, 311

\bibitem[{{Berezhko} \& {V{\"o}lk}(2006)}]{berezhko06}
{Berezhko}, E.~G., \& {V{\"o}lk}, H.~J. 2006, \aap, 451, 981

\bibitem[{{Berezinskii} {et~al.}(1990){Berezinskii}, {Bulanov}, {Dogiel}, \&
  {Ptuskin}}]{berez90}
{Berezinskii}, V.~S., {Bulanov}, S.~V., {Dogiel}, V.~A., \& {Ptuskin}, V.~S.
  1990, {Astrophysics of cosmic rays}, ed. {Berezinskii, V.~S., Bulanov, S.~V.,
  Dogiel, V.~A., \& Ptuskin, V.~S. }

\bibitem[{{Blandford} \& {Eichler}(1987)}]{blandford87}
{Blandford}, R., \& {Eichler}, D. 1987, \physrep, 154, 1

\bibitem[{{Blasi} {et~al.}(2005){Blasi}, {Gabici}, \& {Vannoni}}]{blasi05}
{Blasi}, P., {Gabici}, S., \& {Vannoni}, G. 2005, \mnras, 361, 907

\bibitem[{{Blumenthal} \& {Gould}(1970)}]{blumen70}
{Blumenthal}, G.~R., \& {Gould}, R.~J. 1970, Reviews of Modern Physics, 42, 237

\bibitem[{{Butt} {et~al.}(2008){Butt}, {Porter}, {Katz}, \& {Waxman}}]{butt08}
{Butt}, Y.~M., {Porter}, T.~A., {Katz}, B., \& {Waxman}, E. 2008, \mnras, 386,
  L20

\bibitem[{{Bykov} {et~al.}(2000){Bykov}, {Chevalier}, {Ellison}, \&
  {Uvarov}}]{bykov00}
{Bykov}, A.~M., {Chevalier}, R.~A., {Ellison}, D.~C., \& {Uvarov}, Y.~A. 2000,
  \apj, 538, 203

\bibitem[{{Bykov} {et~al.}(2008){Bykov}, {Uvarov}, \& {Ellison}}]{bykov08}
{Bykov}, A.~M., {Uvarov}, Y.~A., \& {Ellison}, D.~C. 2008, \apjl, 689, L133

\bibitem[{{Cassam-Chena{\"i}} {et~al.}(2004){Cassam-Chena{\"i}},
  {Decourchelle}, {Ballet}, {Sauvageot}, {Dubner}, \& {Giacani}}]{cassam04}
{Cassam-Chena{\"i}}, G., {Decourchelle}, A., {Ballet}, J., {Sauvageot}, J.-L.,
  {Dubner}, G., \& {Giacani}, E. 2004, \aap, 427, 199

\bibitem[{{Chiang} \& {Dermer}(1999)}]{chiang99}
{Chiang}, J., \& {Dermer}, C.~D. 1999, \apj, 512, 699

\bibitem[{{Corless} {et~al.}(1996){Corless}, {Gonnet}, {Hare}, {Jeffrey}, \&
  {Knuth}}]{corless96}
{Corless}, R.~M., {Gonnet}, G.~H., {Hare}, D.~E.~G., {Jeffrey}, D.~J., \&
  {Knuth}, D.~E. 1996, Advances in Computational Mathematics, 5, 329

\bibitem[{{Crusius} \& {Schlickeiser}(1986)}]{crusius86}
{Crusius}, A., \& {Schlickeiser}, R. 1986, \aap, 164, L16

\bibitem[{{CTA Consortium}(2010)}]{cta10}
{CTA Consortium}, T. 2010, arXiv:1008.3703

\bibitem[{{Dermer}(1998)}]{dermer98}
{Dermer}, C.~D. 1998, \apjl, 501, L157

\bibitem[{{Dermer} \& {Menon}(2009)}]{dermer09_book}
{Dermer}, C.~D., \& {Menon}, G. 2009, {High Energy Radiation from Black Holes:
  Gamma Rays, Cosmic Rays, and Neutrinos}

\bibitem[{{Drury} {et~al.}(1994){Drury}, {Aharonian}, \& {Voelk}}]{drury94}
{Drury}, L.~O., {Aharonian}, F.~A., \& {Voelk}, H.~J. 1994, \aap, 287, 959

\bibitem[{{Ellison} {et~al.}(2010){Ellison}, {Patnaude}, {Slane}, \&
  {Raymond}}]{ellison10}
{Ellison}, D.~C., {Patnaude}, D.~J., {Slane}, P., \& {Raymond}, J. 2010, \apj,
  712, 287

\bibitem[{{Enomoto} {et~al.}(2002)}]{enomoto02}
{Enomoto}, R., {et~al.} 2002, \nat, 416, 823

\bibitem[{{Fesen} {et~al.}(2006)}]{fesen06}
{Fesen}, R.~A., {et~al.} 2006, \apj, 645, 283

\bibitem[{{Finke} {et~al.}(2008){Finke}, {Dermer}, \&
  {B{\"o}ttcher}}]{finke08_SSC}
{Finke}, J.~D., {Dermer}, C.~D., \& {B{\"o}ttcher}, M. 2008, \apj, 686, 181

\bibitem[{{Fukui} {et~al.}(2003)}]{fukui03}
{Fukui}, Y., {et~al.} 2003, \pasj, 55, L61

\bibitem[{{Fukui} {et~al.}(2011)}]{fukui11}
---. 2011, \apj, submitted, arXiv:1107.0508

\bibitem[{{Gabici} {et~al.}(2009){Gabici}, {Aharonian}, \&
  {Casanova}}]{gabici09}
{Gabici}, S., {Aharonian}, F.~A., \& {Casanova}, S. 2009, \mnras, 396, 1629

\bibitem[{{Gaisser}(1990)}]{gaisser90}
{Gaisser}, T.~K. 1990, {Cosmic rays and particle physics}, ed. {Gaisser, T.~K.}

\bibitem[{{Ginzburg} \& {Syrovatskii}(1964)}]{ginzburg64}
{Ginzburg}, V.~L., \& {Syrovatskii}, S.~I. 1964, {The Origin of Cosmic Rays},
  ed. {Ginzburg, V.~L.~\& Syrovatskii, S.~I.}

\bibitem[{{Giordano} {et~al.}(2011)}]{giordano11}
{Giordano}, F., {et~al.} 2011, ArXiv e-prints

\bibitem[{{Gould}(1975)}]{gould75}
{Gould}, R.~J. 1975, \apj, 196, 689

\bibitem[{{Green}(2009)}]{green09}
{Green}, D.~A. 2009, Bulletin of the Astronomical Society of India, 37, 45

\bibitem[{{Gupta} {et~al.}(2006){Gupta}, {B{\"o}ttcher}, \& {Dermer}}]{gupta06}
{Gupta}, S., {B{\"o}ttcher}, M., \& {Dermer}, C.~D. 2006, \apj, 644, 409

\bibitem[{{Hayakawa}(1969)}]{hayakawa69}
{Hayakawa}, S. 1969, {Cosmic ray physics. Nuclear and astrophysical aspects},
  ed. {Hayakawa, S.}

\bibitem[{{Hewitt} {et~al.}(2009){Hewitt}, {Yusef-Zadeh}, \&
  {Wardle}}]{hewitt09}
{Hewitt}, J.~W., {Yusef-Zadeh}, F., \& {Wardle}, M. 2009, \apjl, 706, L270

\bibitem[{{Hillas}(2005)}]{hillas05}
{Hillas}, A.~M. 2005, Journal of Physics G Nuclear Physics, 31, 95

\bibitem[{{Inoue} {et~al.}(2011){Inoue}, {Yamazaki}, {Inutsuka}, \&
  {Fukui}}]{inoue11}
{Inoue}, T., {Yamazaki}, R., {Inutsuka}, S.-i., \& {Fukui}, Y. 2011, \apj, in
  press, arXiv:1106.0380

\bibitem[{{Jones}(1968)}]{jones68}
{Jones}, F.~C. 1968, Physical Review, 167, 1159

\bibitem[{{Jones} \& {Ellison}(1991)}]{jones91}
{Jones}, F.~C., \& {Ellison}, D.~C. 1991, \ssr, 58, 259

\bibitem[{{Joshi} \& {B{\"o}ttcher}(2011)}]{joshi11}
{Joshi}, M., \& {B{\"o}ttcher}, M. 2011, \apj, 727, 21

\bibitem[{{Kardashev}(1962)}]{kardashev62}
{Kardashev}, N.~S. 1962, \sovast, 6, 317

\bibitem[{{Katz} \& {Waxman}(2008)}]{katz08}
{Katz}, B., \& {Waxman}, E. 2008, JCAP, 1, 18

\bibitem[{{Kirk}(1994)}]{kirk94}
{Kirk}, J.~G. 1994, in Saas-Fee Advanced Course 24: Plasma Astrophysics, ed.
  {J.~G.~Kirk, D.~B.~Melrose, E.~R.~Priest, A.~O.~Benz, \& T.~J.-L.~Courvoisier
  }, 225

\bibitem[{{Koo} {et~al.}(2004){Koo}, {Kang}, \& {McClure-Griffiths}}]{koo04}
{Koo}, B.-C., {Kang}, J.-H., \& {McClure-Griffiths}, N.~M. 2004, Journal of
  Korean Astronomical Society, 37, 61

\bibitem[{{Krause} {et~al.}(2008){Krause}, {Birkmann}, {Usuda}, {Hattori},
  {Goto}, {Rieke}, \& {Misselt}}]{krause08}
{Krause}, O., {Birkmann}, S.~M., {Usuda}, T., {Hattori}, T., {Goto}, M.,
  {Rieke}, G.~H., \& {Misselt}, K.~A. 2008, Science, 320, 1195

\bibitem[{{Lagage} \& {Cesarsky}(1983)}]{lagage83}
{Lagage}, P.~O., \& {Cesarsky}, C.~J. 1983, \aap, 125, 249

\bibitem[{{Lazendic} {et~al.}(2003){Lazendic}, {Slane}, {Gaensler},
  {Plucinsky}, {Hughes}, {Galloway}, \& {Crawford}}]{lazendic03}
{Lazendic}, J.~S., {Slane}, P.~O., {Gaensler}, B.~M., {Plucinsky}, P.~P.,
  {Hughes}, J.~P., {Galloway}, D.~K., \& {Crawford}, F. 2003, \apjl, 593, L27

\bibitem[{{Lee} {et~al.}(2008){Lee}, {Kamae}, \& {Ellison}}]{lee08}
{Lee}, S., {Kamae}, T., \& {Ellison}, D.~C. 2008, \apj, 686, 325

\bibitem[{{Li} {et~al.}(2011){Li}, {Liu}, \& {Chen}}]{li11}
{Li}, H., {Liu}, S., \& {Chen}, Y. 2011, \apj, in press, arXiv:1110.2857

\bibitem[{{Malkov} \& {Diamond}(2006)}]{malkov06}
{Malkov}, M.~A., \& {Diamond}, P.~H. 2006, \apj, 642, 244

\bibitem[{{Moriguchi} {et~al.}(2005){Moriguchi}, {Tamura}, {Tawara}, {Sasago},
  {Yamaoka}, {Onishi}, \& {Fukui}}]{moriguchi05}
{Moriguchi}, Y., {Tamura}, K., {Tawara}, Y., {Sasago}, H., {Yamaoka}, K.,
  {Onishi}, T., \& {Fukui}, Y. 2005, \apj, 631, 947

\bibitem[{{Morlino} \& {Caprioli}(2011)}]{morlino11}
{Morlino}, G., \& {Caprioli}, D. 2011, ArXiv e-prints

\bibitem[{{Moskalenko} {et~al.}(2006){Moskalenko}, {Porter}, \&
  {Strong}}]{moskalenko06}
{Moskalenko}, I.~V., {Porter}, T.~A., \& {Strong}, A.~W. 2006, \apjl, 640, L155

\bibitem[{{Muraishi} {et~al.}(2000)}]{muraishi00}
{Muraishi}, H., {et~al.} 2000, \aap, 354, L57

\bibitem[{{Porter} {et~al.}(2006){Porter}, {Moskalenko}, \&
  {Strong}}]{porter06}
{Porter}, T.~A., {Moskalenko}, I.~V., \& {Strong}, A.~W. 2006, \apjl, 648, L29

\bibitem[{{Reimer} \& {Pohl}(2002)}]{reimer02}
{Reimer}, O., \& {Pohl}, M. 2002, \aap, 390, L43

\bibitem[{{Reynolds}(1998)}]{reynolds98}
{Reynolds}, S.~P. 1998, \apj, 493, 375

\bibitem[{{Reynolds}(2008)}]{reynolds08}
---. 2008, \araa, 46, 89

\bibitem[{{Reynolds}(2010)}]{reynolds10}
---. 2010, \apss, 407

\bibitem[{{Reynolds} {et~al.}(2011){Reynolds}, {Gaensler}, \&
  {Bocchino}}]{reynolds11}
{Reynolds}, S.~P., {Gaensler}, B.~M., \& {Bocchino}, F. 2011, \ssr, 269

\bibitem[{{Slane} {et~al.}(1999){Slane}, {Gaensler}, {Dame}, {Hughes},
  {Plucinsky}, \& {Green}}]{slane99}
{Slane}, P., {Gaensler}, B.~M., {Dame}, T.~M., {Hughes}, J.~P., {Plucinsky},
  P.~P., \& {Green}, A. 1999, \apj, 525, 357

\bibitem[{{Slane} {et~al.}(2002){Slane}, {Smith}, {Hughes}, \&
  {Petre}}]{slane02}
{Slane}, P., {Smith}, R.~K., {Hughes}, J.~P., \& {Petre}, R. 2002, \apj, 564,
  284

\bibitem[{{Sturner} {et~al.}(1997){Sturner}, {Skibo}, {Dermer}, \&
  {Mattox}}]{sturner97}
{Sturner}, S.~J., {Skibo}, J.~G., {Dermer}, C.~D., \& {Mattox}, J.~R. 1997,
  \apj, 490, 619

\bibitem[{{Tanaka} {et~al.}(2008)}]{tanaka08}
{Tanaka}, T., {et~al.} 2008, \apj, 685, 988

\bibitem[{{Telezhinsky} {et~al.}(2012){Telezhinsky}, {Dwarkadas}, \&
  {Pohl}}]{telez12}
{Telezhinsky}, I., {Dwarkadas}, V.~V., \& {Pohl}, M. 2012, Astroparticle
  Physics, 35, 300

\bibitem[{{Truelove} \& {McKee}(1999)}]{truelove99}
{Truelove}, J.~K., \& {McKee}, C.~F. 1999, \apjs, 120, 299

\bibitem[{{Uchiyama} \& {Aharonian}(2008)}]{uchiy08}
{Uchiyama}, Y., \& {Aharonian}, F.~A. 2008, \apjl, 677, L105

\bibitem[{{Uchiyama} {et~al.}(2003){Uchiyama}, {Aharonian}, \&
  {Takahashi}}]{uchiy03}
{Uchiyama}, Y., {Aharonian}, F.~A., \& {Takahashi}, T. 2003, \aap, 400, 567

\bibitem[{{Uchiyama} {et~al.}(2007){Uchiyama}, {Aharonian}, {Tanaka},
  {Takahashi}, \& {Maeda}}]{uchiy07}
{Uchiyama}, Y., {Aharonian}, F.~A., {Tanaka}, T., {Takahashi}, T., \& {Maeda},
  Y. 2007, \nat, 449, 576

\bibitem[{{Uchiyama} {et~al.}(2010){Uchiyama}, {Blandford}, {Funk}, {Tajima},
  \& {Tanaka}}]{uchiyama10}
{Uchiyama}, Y., {Blandford}, R.~D., {Funk}, S., {Tajima}, H., \& {Tanaka}, T.
  2010, \apjl, 723, L122

\bibitem[{{Wang} {et~al.}(1997){Wang}, {Qu}, \& {Chen}}]{wang97}
{Wang}, Z.~R., {Qu}, Q., \& {Chen}, Y. 1997, \aap, 318, L59

\bibitem[{{Yamazaki} {et~al.}(2009){Yamazaki}, {Kohri}, \&
  {Katagiri}}]{yamazaki09}
{Yamazaki}, R., {Kohri}, K., \& {Katagiri}, H. 2009, \aap, 495, 9

\bibitem[{{Yuan} {et~al.}(2011{\natexlab{a}}){Yuan}, {Liu}, {Fan}, {Bi}, \&
  {Fryer}}]{yuan11}
{Yuan}, Q., {Liu}, S., {Fan}, Z., {Bi}, X., \& {Fryer}, C.~L.
  2011{\natexlab{a}}, \apj, 735, 120

\bibitem[{{Yuan} {et~al.}(2011{\natexlab{b}}){Yuan}, {Yin}, \&
  {Bi}}]{yuan11_neutrino}
{Yuan}, Q., {Yin}, P.-F., \& {Bi}, X.-J. 2011{\natexlab{b}}, Astroparticle
  Physics, 35, 33

\bibitem[{{Zirakashvili} \& {Aharonian}(2007)}]{zirak07}
{Zirakashvili}, V.~N., \& {Aharonian}, F. 2007, \aap, 465, 695

\bibitem[{{Zirakashvili} \& {Aharonian}(2010)}]{zirak10}
{Zirakashvili}, V.~N., \& {Aharonian}, F.~A. 2010, \apj, 708, 965

\end{thebibliography}

%\end{thebibliography}

\end{document}